\documentstyle{article}
\newcommand{\wsj}[6]{\left\{ \begin{array}{rrr}
                             #1 & #2 & #3 \\
                             #4 & #5 & #6 
                             \end{array} \right\} }
\newcommand{\nph}[1]{#1$p$#1$h$}

\begin{document}

\title{Ground State Correlations and Mean Field Using the exp(S) Method.}
\author{Jochen H. Heisenberg and Bogdan Mihaila \\
        Department of Physics, University of New Hampshire, 
        Durham, New Hampshire 03824}
\maketitle

\abstract{
This document\footnote{work supported by DOE grant DE-FG02-87ER-40371}
gives a detailed account of the terms
used in the computation of the ground state mean field and the
ground state correlations. While the general approach to this
description is given in a separate paper (nucl-th/9802029)
we give here the explicite expressions used.
}



\tableofcontents

\newpage
\section{Preliminaries.}
\label{sec:I}

The purpose of this document is to give a detailed account of the
terms which we use in generating the 2-body ground state correlations.
The terms are expectation values of operator products in the
bare ground state.
In order to correct for the presence of \nph{3} and \nph{4} correlations,
the operator products are generated via a $1/\epsilon$ expansion,
as outlined in reference \cite{ref:paper_one}.
As such, this document can be considered as an appendix to that paper.

Expressions here are given with and without angular momentum coupling (AMC).
The calculations were done in angular momentum coupling
and required considerable recoupling from ($ph$) coupling
to ($pp$) coupling and vice versa. ($ph$) coupling is indicated
by means of small or greek letters ($\ell$,$\lambda$);
whereas ($pp$) coupling is indicated by capital letters ($L$,$K$).
The phase-conventions of our matrix elements and amplitudes leading
to their symmetries as well as the details of the computation
can be found in reference \cite{ref:2b_me}.

The terms can be classified as either contributing to the effective
two-body matrix element or contributing to the effective single
particle energy.
The contributions for the effective
matrix element are $B^{eff}(p_1h_1,p_2h_2)=
\langle p_1\bar h_1| V^{eff,\lambda}| h_2\bar p_2\rangle$.
If this matrix element is given in ($pp$) coupling, an extra recoupling
is implied according to
\begin{eqnarray}
   \langle p_1\bar h_1| V^{eff,\lambda}| h_2\bar p_2\rangle=
   \sum_K(-)^{K+1} \, (2K+1) \
       \wsj{p_1}{h_1}{\lambda}{h_2}{p_2}{K} \
   \langle p_1p_2 | V^{K,eff} | h_1h_2\rangle
\end{eqnarray}
We are using Einstein summation convention,
so that any orbit appearing twice implies summation over that orbit.

In most cases we deal with the energy ($\omega$) dependence of the
two-body matrix element the following way:
\begin{eqnarray}
   \sum_n \frac{b_n}{\epsilon_n+\omega}\approx
   \Bigl( \sum_n \frac{b_n}{\epsilon_n} \Bigr)
   \Bigl( \frac{1}{1+\alpha\omega} \Bigr)
\end{eqnarray}
with
\begin{eqnarray}
   \alpha=\Bigl(\sum_n \frac{b_n}{\epsilon_n^2} \Bigr) 
   /\Bigl(\sum_n \frac{b_n}{\epsilon_n} \Bigr)
\end{eqnarray}
This expression gives the correct value at $\omega=0$ and
$\omega=\infty$, and it also
gives the correct derivative with respect to $\omega$ at ($\omega=0$).
When $\omega$ dependent single-particle energies are required,
we compute these energies for a set of 9 values of $\omega$
in the range from 0 to 1 GeV, and interpolate for values inbetween.

We also define the density matrix
\begin{eqnarray}
   d(p_1,p_3,\omega)
   & = & 
   - \frac{1}{2} \ 
     \frac{Z_{p_3p_2,h_4h_2}
           V_{p_1p_2,h_4h_2}}
          {\epsilon_{p_2h_2}+\epsilon_{p_1h_4}+\omega}
   \\
   d(h_1,h_3,\omega)
   & = & 
   - \frac{1}{2} \
     \frac{Z_{p_4p_2,h_3h_2}
           V_{p_4p_2,h_1h_2}}
          {\epsilon_{p_2h_2}+\epsilon_{p_4h_1}+\omega}
\end{eqnarray}
or, in angular momentum coupling
\begin{eqnarray}
   &&
   d(p_1,p_3,\omega)=
   -\frac{1}{2} \sum_{\ell} \frac{2\ell+1}{2j_{p_1}+1}
   Z^{\ell}_{p_1h_1,p_2h_2}
   \frac{\langle p_2 \bar h_2 | V^{\ell} | h_1 \bar p_3\rangle}
        {\epsilon_{p_3h_1}+\epsilon_{p_2h_2}+\omega}
   \\ &&
   d(h_1,h_3,\omega)=
   -\frac{1}{2} \sum_{\ell} \frac{2\ell+1}{2j_{h_1}+1}
   Z^{\ell}_{p_1h_1,p_2h_2}
   \frac{\langle p_2 \bar h_2 | V^{\ell} | h_1 \bar p_3\rangle}
        {\epsilon_{p_3h_1}+\epsilon_{p_2h_2}+\omega}
\end{eqnarray}

\section{Mean Field $ph$ Matrix elements.}
\label{sec:II}

We evaluate the terms
\begin{eqnarray}
   \langle 0 | U | ph \rangle \ = \ 
   \langle 0|{\bf V}_{01}|ph\rangle +\langle 0|[{\bf S}_2,{\bf V}_{10}]|ph\rangle 
   +\langle 0|[{\bf S}_3,{\bf V}_{20}]|ph\rangle 
\end{eqnarray}
The ${\bf S}_3$ term requires a $1/\epsilon$ expansion
\begin{eqnarray}
   &&
   -\langle 0|\Bigl[\bigl[{\bf S}_2,{\mathbf V}_{01}\bigr],{\mathbf V}_{20}\Bigr]|ph
   \rangle/\epsilon
   -\frac{1}{2} \langle 0|\Bigl[{\bf S}_2,\bigl[{\bf S}_2,{\mathbf V}_{10}\bigr]\Bigr]
   {\mathbf V}_{20}|ph\rangle/\epsilon
   \\ &&
   -\frac{1}{\epsilon_3}
   \langle 0|\bigl[{\bf S}_3,{\mathbf V}_{00}\bigr],{\mathbf V}_{20}|ph\rangle
   -\frac{1}{\epsilon_3}
   \langle 0|\bigl[{\bf S}_4,{\mathbf V}_{10}\bigr],{\mathbf V}_{20}|ph\rangle
   -\frac{1}{\epsilon_3}
   \langle 0|\bigl[{\bf S}_5,{\mathbf V}_{20}\bigr],{\mathbf V}_{20}|ph\rangle
   -\frac{1}{\epsilon_3}
   \langle 0|\Bigl[{\bf S}_3,\bigl[{\bf S}_2,{\mathbf V}_{20}\bigr]\Bigr]
   {\mathbf V}_{20}|ph\rangle
   \nonumber \\
\end{eqnarray}
Up to next order, we can write the terms in the last line (keeping only
terms with ${\bf S}_2$) as
\begin{eqnarray}
   &&
   +\frac{1}{\epsilon_3}\frac{1}{\epsilon_3'}
   \langle 0|\Bigl[\bigl[{\bf S}_2,{\mathbf V}_{01}\bigr],{\mathbf V}_{00}\Bigr],
   {\mathbf V}_{20}|ph\rangle
   +\frac{1}{\epsilon_3}\frac{1}{\epsilon_3'}\frac{1}{2} 
   \langle 0|\biggl[\Bigl[{\bf S}_2,\bigl[{\bf S}_2,{\mathbf V}_{10}\bigr]
   \Bigr],{\mathbf V}_{00}\biggr],{\mathbf V}_{20}|ph\rangle
   \\ &&
   +\frac{1}{\epsilon_3}\frac{1}{\epsilon_4}\frac{1}{2} 
   \langle 0|\Bigl[{\bf S}_2,\bigl[{\bf S}_2,{\mathbf V}_{00}\bigr]
   \Bigr],{\mathbf V}_{10}{\mathbf V}_{20}|ph\rangle
   +\frac{1}{\epsilon_3}\frac{1}{\epsilon_4}\frac{1}{6} 
   \langle 0|\Bigl[{\bf S}_2,\bigl[{\bf S}_2,[{\bf S}_2,{\mathbf V}_{20}]\bigr]
   \Bigr],{\mathbf V}_{10}{\mathbf V}_{20}|ph\rangle
   \\ &&
   +\frac{1}{\epsilon_3}\frac{1}{\epsilon_5}\frac{1}{6} 
   \langle 0|\Bigl[{\bf S}_2,\bigl[{\bf S}_2,[{\bf S}_2,{\mathbf V}_{10}]\bigr]
   \Bigr],{\mathbf V}_{20}{\mathbf V}_{20}|ph\rangle
   \\ &&
   +\frac{1}{\epsilon_3}\frac{1}{\epsilon_3'}
   \langle 0|\Bigl[[{\bf S}_2,{\mathbf V}_{01}],\bigl[{\bf S}_2,{\mathbf V}_{20}
   \bigr]\Bigr] {\mathbf V}_{20}|ph\rangle
   +\frac{1}{\epsilon_3} \frac{1}{\epsilon_3'}
   \langle 0|\Bigl[\bigl[{\bf S}_2,[{\bf S}_2,{\mathbf V}_{10}]\bigr],
   \bigl[{\bf S}_2,{\mathbf V}_{20}
   \bigr]\Bigr] {\mathbf V}_{20}|ph\rangle
\end{eqnarray}

\paragraph{Term 1: $\langle 0| {\mathbf V}_{01} |ph\rangle$.}
\begin{eqnarray}
   (-)^{j_p-j_{h_s}} \,
   \sqrt{ \frac{2j_{h_s}+1}{2j_p+1} } \langle p \bar h| V^{\ell=0}
   | h_s \bar h_s\rangle
\end{eqnarray}
Term (1) is modified by terms (3b) and (3e). 
These terms are calculated by finding a representation of natural orbits
$\lbrace p^n,h^n\rbrace=\lbrace o_n\rbrace$,
in which the density matrix is diagonal with a diagonal value $v_o$, 
the occupation of the natural orbits. 
Thus these terms can be combined as
\begin{eqnarray}
   \sum_{o_n}(-)^{k_p+k_{o_n}} \, 
   \sqrt{ \frac{2j_{o_n}+1}{2j_p+1} } 
   \langle p {\bar h} | V^{\ell=0} | o_n {\bar o}_n \rangle v_{o_n}
\end{eqnarray}

\paragraph{Term 2: $\langle 0| [{\bf S}_2,{\bf V}_{10}] |ph\rangle$.}
\begin{eqnarray}
   &&
   (a)\quad +Z_{p_1h_1,p_2h}\langle p_1h_1|V|pp_2\rangle 
      \quad \rightarrow \quad
      \frac{1}{2}  \sum_{\ell} \frac{2\ell+1}{2j_p+1}
      Z^{\ell}_{p_1h,p_2h_2}~\langle p_1 \bar p | V^{\ell} | h_2 \bar p_2\rangle
   \\ &&
   (b)\quad -Z_{p_1h_1,ph_2}\langle p_1h_1|V|h_2h\rangle 
      \quad \rightarrow \quad
      -\frac{1}{2}  \sum_{\ell} \frac{2\ell+1}{2j_p+1}
      Z^{\ell}_{ph_1,p_2h_2}~\langle h \bar h_1 | V^{\ell} | h_2 \bar p_2\rangle
   \\ &&
   (c)\quad +Z^0_{ph,p_1h_1}\langle p_1h_1|V|h_sh_s\rangle 
      \quad \rightarrow \quad
      Z^0_{ph,p_2h_2}~\langle p_2| H| h_2\rangle
\end{eqnarray}
Here $H$ represents the one-body part of the hamiltonian. As for
each term there appears also a term multiplying it by $Z^0$, the sum of
these terms is small in the hartree-fock basis as the one-body part
already is small.

\paragraph{Term 3: $\langle 0|
                        \bigl [
                           [{\bf S}_2,{\bf V}_{01}], \frac{1}{\bf H_0}{\bf V}_{20}
                        \bigr ]
                    |ph\rangle$.}
\begin{enumerate}
   \item{Term 3a.}
        \begin{eqnarray}
           &&
           -Z_{p_1h,p_2h_2}V_{p_3ph_3p_1}V_{p_2p_3h_2h_3}/\epsilon
           \nonumber \\ &&
           \rightarrow \quad
           -\sum_{\ell} \frac{2\ell+1}{2j_p+1} Z^{\ell}_{p_1h,p_2h_2}
           \langle p_3 \bar h_3| V^{\ell}| p_1 \bar p\rangle
           \frac{\langle p_2 \bar h_2| V^{\ell}| h_3 \bar p_3\rangle}
                {\epsilon_{ph}+\epsilon_{p_2h_2}+\epsilon_{p_3h_3}}
        \end{eqnarray}
   \item{Term 3b.}
        \begin{eqnarray}
           &&
           -\frac{1}{2} Z_{p_1h_1,p_2h_2}V_{pp_3hp_1}V_{p_2p_3h_2h_1}/\epsilon
           \nonumber \\ &&
           \rightarrow \quad
           (-)^{j_p-j_{p_1}} \sqrt{ \frac{2j_{p_1}+1}{2j_p+1} }
           \langle p \bar h| V^{\ell=0}| p_1 \bar p_3\rangle \ 
           d(p_1,p_3,\omega=\epsilon_{ph})
        \end{eqnarray}
   \item{Term 3c.}
        \begin{eqnarray}
           &&
           +\frac{1}{4} Z_{ph_1,p_2h_2}V_{p_1p_3hp_2}V_{p_2p_3h_2h_1}/\epsilon
           \nonumber \\ &&
           \rightarrow \quad
           \frac{1}{4} \sum_K \frac{2K+1}{2j_p+1} Z^K_{pp_2,h_1h_2}
           \langle p_1p_3| V^K| hp_2\rangle
           \frac{\langle p_1p_3| V^K| h_1h_2\rangle}
                {\epsilon_{ph}+\epsilon_{p_1h_1}+\epsilon_{p_2h_2}}
        \end{eqnarray}
   \item{Term 3d.}
        \begin{eqnarray}
           &&
           Z_{ph_1,p_2h_2}V_{p_3h_1h_3h}V_{p_2p_3h_2h_3}/\epsilon
           \nonumber \\ &&
           \rightarrow \quad
           - \frac{2\ell+1}{2j_p+1}
           Z^{\ell}_{ph_1,p_2h_2}Z^{\ell}_{p_2h_2,p_3h_3}
           \langle p_3 \bar h_3 | V^{\ell} | h \bar h_1\rangle
        \end{eqnarray}
   \item{Term 3e.}
        \begin{eqnarray}
           &&
           \frac{1}{2} Z_{p_1h_1,p_2h_2}V_{ph_2hh_3}V_{p_1p_2h_1h_3}/\epsilon
           \nonumber \\ &&
           \rightarrow \quad
           -(-)^{j_p-j_{h_2}} \ \sqrt{ \frac{2j_{h_2}+1}{2j_p+1} }
           \langle p \bar h | V^{\ell=0} | h_3 \bar h_2\rangle
           \sum_{\ell} -\frac{1}{2} \frac{2\ell+1}{2j_{h_2}+1} Z^{\ell}_{p_1h_1,p_2h_2}
           \frac{\langle p_1 \bar h_1 | V^{\ell} | h_3 \bar p_2\rangle}
                {\epsilon_{ph}+\epsilon_{p_1h_1}+\epsilon_{p_2h_3}}
        \end{eqnarray}
   \item{Term 3f.}
        \begin{eqnarray}
           &&
           -\frac{1}{4} Z_{p_1h,p_2h_2}V_{ph_2h_3h_1}V_{p_1p_2h_3h_1}/\epsilon
           \nonumber \\ &&
           \rightarrow \quad
           -\frac{1}{4} \sum_K \frac{2K+1}{2j_p+1} Z^K_{p_1p_2,hh_2}
           \langle ph_2| V^K| h_3h_1\rangle
           \frac{\langle p_1p_2| V^K| h_3h_1\rangle}
                {\epsilon_{ph}+\epsilon_{p_1h_1}+\epsilon_{p_2h_3}}
        \end{eqnarray}
   \item{Term 3g.}
        \begin{eqnarray}
           &&
           -\frac{1}{2} Z_{p_1h_1,ph}V_{p_3p_2h_3p_1}V_{p_2p_3h_1h_3}/\epsilon
           \nonumber \\ &&
           \rightarrow \quad
           -(-)^{k_p+k_{h_1}}\sqrt{ \frac{2j_{h_1}+1}{2j_p+1} } \ 
           Z^0_{ph,p_1h_1} \sum_\ell \frac{2\ell+1}{2j_{h_1}+1} \langle p_3h_3|V^{\ell}|p_1p_2\rangle  
           \frac{B^{\ell}_{p_3h_3,p_2h_1}}
                {\epsilon_{p_3h_3}+\epsilon_{p_2h_1}+\epsilon_{ph}}
        \end{eqnarray}
   \item{Term 3h.}
        \begin{eqnarray}
           &&
           \frac{1}{2} Z_{p_1h_1,ph}V_{h_2h_3h_1p_3}V_{p_1p_3h_2h_3}/\epsilon
           \nonumber \\ &&
           \rightarrow \quad
           (-)^{k_p+k_{h_1}}\sqrt{ \frac{2j_{h_1}+1}{2j_p+1} } \ 
           Z^0_{ph,p_1h_1}
           \sum_\ell \frac{2\ell+1}{2j_{h_1}+1} \langle p_3h_3|V^{\ell}|h_2h_1\rangle 
           \frac{\langle p_1p_2| V^K| h_3h_4\rangle}
                {\epsilon_{ph}+\epsilon_{p_1h_3}+\epsilon_{p_2h_4}}
        \end{eqnarray}
\end{enumerate}

\paragraph{Term 4: $\langle 0|
                        \Bigl [
                           \bigl [ 
                              {\bf S}_2, [{\bf S}_2,{\bf V}_{10}] 
                           \bigr ], 
                           \frac{1}{\bf H_0}{\bf V}_{20}
                        \Bigr ]
                    |ph\rangle$.}
\begin{enumerate}
   \item{Term 4a.}
        \begin{eqnarray}
           &&
           -\frac{1}{8} Z_{p_1h,p_2h_2}Z_{p_3h_3,p_4h_4}V_{ph_2p_3p_4}V_{p_1p_2h_3h_4}/\epsilon
           \nonumber \\ &&
           \rightarrow \quad
           -\frac{1}{8} \sum_K \frac{2K+1}{2j_p+1} Z^K_{p_1p_2,hh_2}Z^K_{p_3p_4,h_3h_4}
           \langle p_3p_4| V^K| ph_2\rangle
           \frac{\langle p_1p_2| V^K| h_3h_4\rangle}
                {\epsilon_{ph}+\epsilon_{p_1h_3}+\epsilon_{p_2h_4}}
        \end{eqnarray}   
   \item{Term 4b.}
        \begin{eqnarray}
           &&
           +\frac{1}{4} Z_{p_1h_1,p_2h_2}Z_{p_3h,p_4h_4}V_{ph_2p_3p_4}V_{p_1p_2h_1h_4}/\epsilon
           \nonumber \\ &&
           \rightarrow \quad
           -\frac{1}{2} \sum_{\ell} \frac{2\ell+1}{2j_p+1}
           Z^{\ell}_{p_1h,p_2h_2}\langle h_3 \bar p_2| V^{\ell}| p_1 \bar p\rangle \
           d(h_3,h_2,\omega=\epsilon_{ph})
        \end{eqnarray}
   \item{Term 4c.}
        \begin{eqnarray}
           &&
           +\frac{1}{4} Z_{p_1h,p_2h_2}Z_{p_3h_3,p_4h_4}V_{ph_2p_3p_2}V_{p_1p_4h_3h_4}/\epsilon
           \nonumber \\ &&
           \rightarrow \quad
           -\frac{1}{2} \sum_{\ell} \frac{2\ell+1}{2j_p+1}
           Z^{\ell}_{p_1h,p_2h_2}\langle h_2 \bar p_2| V^{\ell}| p_3 \bar p\rangle \
           d(p_3,p_1,\omega=\epsilon_{ph})
        \end{eqnarray}
   \item{Term 4d.}
        \begin{eqnarray}
           &&
           -\frac{1}{2} Z_{p_1h_1,p_2h_2}Z_{p_3h,p_4h_4}V_{ph_2p_3p_2}V_{p_1p_4h_1h_4}/\epsilon
           \nonumber \\ &&
           \rightarrow \quad
           -\frac{1}{2} \sum_{\ell} \frac{2\ell+1}{2j_p+1}
           Z^{\ell}_{p_1h_1,p_2h_2}Z^{\ell}_{p_3h,p_4h_4}
           \langle p_3 \bar p| V^{\ell}| h_2 \bar p_2\rangle
           \frac{\langle p_1 \bar h_1| V^{\ell}| h_4 \bar p_4\rangle}
                {\epsilon_{ph}+\epsilon_{p_1h_1}+\epsilon_{p_4h_4}}
        \end{eqnarray}
   \item{Term 4e.}
        \begin{eqnarray}
           &&
           -\frac{1}{4} Z_{p_1h_1,p_2h_2}Z_{ph_2,p_3h_3}V_{h_3h_4,p_3h}V_{p_1p_2h_1h_4}/\epsilon
           \nonumber \\ &&
           \rightarrow \quad
           +\frac{1}{2} \sum_{\ell} \frac{2\ell+1}{2j_p+1}
           Z^{\ell}_{ph_1,p_2h_2}\langle h_2 \bar p_2| V^{\ell}| h \bar h_3\rangle \
           d(h_3,h_1,\omega=\epsilon_{ph})
        \end{eqnarray}
   \item{Term 4f.}
        \begin{eqnarray}
           &&
           -\frac{1}{4} Z_{ph_1,p_2h_2}Z_{p_3h_3,p_4h_4}V_{h_1h_2,hp_3}V_{p_1p_4h_3h_4}/\epsilon
           \nonumber \\ &&
           \rightarrow \quad
           +\frac{1}{2} \sum_{\ell} \frac{2\ell+1}{2j_p+1}
           Z^{\ell}_{ph_1,p_2h_2}\langle h_2 \bar p_3| V^{\ell}| h \bar h_1\rangle \
           d(p_3,p_2,\omega=\epsilon_{ph})
        \end{eqnarray}
   \item{Term 4j.}
        \begin{eqnarray}
           &&
           Z_{ph_1,p_2h_2}Z_{p_3h,p_4h_4}V_{p_5h_2p_3p_2}V_{p_5p_4h_1h_4}/\epsilon
           \nonumber \\ &&
           \rightarrow \quad
           \sum_{\ell} \frac{2\ell+1}{2j_p+1}
           Z^{\ell}_{ph_1,p_2h_2}Z^{\ell}_{p_3h,p_4h_4}
           \langle p_2 \bar h_2 | V^{\ell}| p_5 \bar p_3\rangle
           \frac{\langle p_5 \bar h_1| V^{\ell}| h_4 \bar p_4\rangle}
                {\epsilon_{ph}+\epsilon_{p_5h_1}+\epsilon_{p_4h_4}}
        \end{eqnarray}
   \item{Term 4k.}
        \begin{eqnarray}
           &&
           \frac{1}{8} Z_{ph_1,p_2h_2}Z_{p_3h_3,p_4h_4}V_{hp_2h_3h_4}V_{p_3p_4h_1h_2}/\epsilon
           \nonumber \\ &&
           \rightarrow \quad
           \frac{1}{8} \sum_K \frac{2K+1}{2j_p+1} Z^K_{pp_2,h_1h_2}Z^K_{p_3p_4,h_3h_4}
           \langle h_3h_4| V^K| hp_2\rangle
           \frac{\langle h_1h_2| V^K| p_3p_4\rangle}
                {\epsilon_{ph}+\epsilon_{p_3h_1}+\epsilon_{p_4h_2}}
        \end{eqnarray}
   \item{Term 4l.}
        \begin{eqnarray}
           &&
           +\frac{1}{2} Z_{p_1h_1,p_2h_2}Z_{ph_3,p_4h_4}V_{h_3h_2hp_2}V_{p_1p_4h_1h_4}/\epsilon
           \nonumber \\ &&
           \rightarrow \quad
           +\frac{1}{2} \sum_{\ell} \frac{2\ell+1}{2j_p+1}
           Z^{\ell}_{p_1h_1,p_2h_2}Z^{\ell}_{ph_3,p_4h_4}
           \langle h_3 \bar h| V^{\ell}| p_2 \bar h_2\rangle
           \frac{\langle p_1 \bar h_1| V^{\ell}| h_4 \bar p_4\rangle}
                {\epsilon_{ph}+\epsilon_{p_1h_1}+\epsilon_{p_4h_4}}
        \end{eqnarray}
   \item{Term 4o.}
        \begin{eqnarray}
           &&
           -Z_{p_1h_1,p_2h}Z_{ph_3,p_4h_4}V_{h_5p_1h_3h_1}V_{p_1p_4h_5h_4}/\epsilon
           \nonumber \\ &&
           \rightarrow \quad
           -\sum_{\ell} \frac{2\ell+1}{2j_p+1}
           Z^{\ell}_{p_1h_1,p_2h}Z^{\ell}_{ph_3,p_4h_4}
           \langle p_1 \bar h_1 | V^{\ell} | h_3 \bar h_5\rangle
           \frac{\langle p_1 \bar h_5| V^{\ell}| h_4 \bar p_4\rangle}
                {\epsilon_{ph}+\epsilon_{p_1h_5}+\epsilon_{p_4h_4}}
        \end{eqnarray}
\end{enumerate}

\paragraph{Term 5: $\langle 0|
                        \Bigl [
                           \bigl [
                              [{\bf S}_2,{\bf V}_{01}], \frac{1}{\bf H_0}{\bf V}_{00}
                           \bigr ], 
                           \frac{1}{\bf H_0}{\bf V}_{20}
                        \Bigr ]
                    |ph\rangle$.}
\begin{enumerate}
   \item{Term 5a.}
        \begin{eqnarray}
           \sum_l \ \frac{2l+1}{2j_p+1} \ 
           \frac{Z^l{p_1h,p_2h_2}
                 \langle p_1 \bar h_2 | V^l | h_3 \bar p_3 \rangle
                 \langle h_3 \bar p_3 | V^l | h_4 \bar p_4 \rangle
                 \langle h_4 \bar p_4 | V^l | p_1 \bar p   \rangle }
                {(\epsilon_{p_2h_2}+\epsilon_{p_3h_3}+\epsilon_{ph})
                 (\epsilon_{p_4h_4}+\epsilon_{p_2h_2}+\epsilon_{ph}) }
        \end{eqnarray}
   \item{Term 5b.}
        \begin{eqnarray}
           -
           \sum_l \ \frac{2l+1}{2j_p+1} \ 
           \frac{Z^l{ph_1,p_2h_2}
                 \langle p_1 \bar h_2 | V^l | h_3 \bar p_3 \rangle
                 \langle h_3 \bar p_3 | V^l | h_4 \bar p_4 \rangle
                 \langle h_4 \bar p_4 | V^l | h \bar h_1   \rangle }
                {(\epsilon_{p_2h_2}+\epsilon_{p_3h_3}+\epsilon_{ph})
                 (\epsilon_{p_4h_4}+\epsilon_{p_2h_2}+\epsilon_{ph}) }
        \end{eqnarray}
\end{enumerate}

\section{Z Coefficients.}
\label{sec:III}

All contributions are identified by a Roman numeral indicating the
operator from the $1/\epsilon$ expansion that is generating this particular term
according to the following list:

\begin{tabular}{ll}
   $
   \langle 0 | {\mathbf V}_{02} | 2p2h\rangle  
   + \langle 0 | \big[ {\bf S}_2, {\mathbf V}_{00} \big] | 2p2h\rangle  
   + \frac{1}{2} \langle  0 | \Big[ 
                              {\bf S}_2, \big[ {\bf S}_2, {\mathbf V}_{20} \big]
                           \Big] | 2p2h\rangle   
   $
   & (I,II,III)
   \\ 
   $
   - \frac{1}{\epsilon_{3p3h}}
     \langle 0 | \Big[ \big[ {\bf S}_2, {\mathbf V}_{01}\big], 
                    {\mathbf V}_{10} 
              \Big] | 2p2h\rangle 
   - \frac{1}{2} \frac{1}{\epsilon_{3p3h}}
     \langle 0 | \bigg[ \Big[ {\bf S}_2,
                           \big[{\bf S}_2, {\mathbf V}_{10} \big] 
                     \Big], {\mathbf V}_{10}
              \bigg] | 2p2h\rangle 
   $
   & (IV,V)
   \\  
   $
   - \frac{1}{2} \frac{1}{\epsilon_{4p4h}}
     \langle  0 | \bigg[ \Big[ {\bf S}_2, \big[ {\bf S}_2, {\mathbf V}_{00} \big]
                      \Big], {\mathbf V}_{20}
               \bigg] | 2p2h\rangle 
   + \frac{1}{\epsilon_{3p3h}} \frac{1}{\epsilon_{3p3h}}
     \langle 0 | \bigg[ \Big[ \big[ {\bf S}_2, {\mathbf V}_{01} \big], 
                           {\mathbf V}_{00}
                     \Big], {\mathbf V}_{10}
              \bigg] | 2p2h\rangle 
   $
   & (VI,VII)
   \\ 
   $
   + \frac{1}{2} \frac{1}{\epsilon_{4p4h}} \frac{1}{\epsilon_{3p3h}}
     \langle 0 | \bigg[ \Big[ \big[{\bf S}_2, {\mathbf V}_{01} \big], 
                           {\mathbf V}_{01}
                     \Big], {\mathbf V}_{20}
              \bigg] | 2p2h\rangle 
   $
   & (VIII)
   \\ 
   $
   - \frac{1}{6} \frac{1}{\epsilon_{4p4h}}
     \langle 0 | \Bigg[ \bigg[ {\bf S}_2, \Big[ {\bf S}_2, \big[{\bf S}_2, {\mathbf V}_{20}\big]
                                       \Big]
                     \bigg], {\mathbf V}_{20} 
              \Bigg] | 2p2h\rangle 
   $
   & (IX)
   \\ 
   $
   + \frac{1}{\epsilon_{3p3h}} \frac{1}{\epsilon_{3p3h}}
     \langle 0 | \bigg[ \Big[ \big[ {\bf S}_2, {\mathbf V}_{01} \big],
                           \big[ {\bf S}_2, {\mathbf V}_{20} \big]
                     \Big],
                     {\mathbf V}_{10} 
              \bigg] | 2p2h\rangle 
   $
   & (X)
   \\ 
   $
   + \frac{1}{\epsilon_{4p4h}} \frac{1}{\epsilon_{3p3h}}
     \langle 0 | \bigg[ \Big[ \big[{\bf S}_2, {\mathbf V}_{01}\big], 
                           \big[{\bf S}_2,{\mathbf V}_{10}\big]
                     \Big], {\mathbf V}_{20}
              \bigg] | 2p2h\rangle 
   $
   & (XI)
   \\ 
   $
   - \frac{1}{\epsilon_{3p3h}} \frac{1}{\epsilon_{4p4h}} \frac{1}{\epsilon_{3p3h}}
     \langle 0 | \Bigg[ \bigg[ \Big[ \big[ {\bf S}_2, {\mathbf V}_{01} \big],
                                  {\mathbf V}_{01}
                            \Big], {\mathbf V}_{10}
                     \bigg], {\mathbf V}_{10}
              \Bigg] | 2p2h\rangle 
   $
   & (XII)
   \\ 
   $
   - \frac{1}{\epsilon_{4p4h}} \frac{1}{\epsilon_{4p4h}} \frac{1}{\epsilon_{3p3h}}
     \langle 0 | \Bigg[ \bigg[ \Big[ \big[ {\bf S}_2, {\mathbf V}_{01} \big],
                                  {\mathbf V}_{01} 
                            \Big], {\mathbf V}_{00}
                     \bigg], {\mathbf V}_{20}
              \Bigg] | 2p2h\rangle 
   $
   & (XIII)
   \\ 
   $
   - \frac{1}{\epsilon_{3p3h}} \frac{1}{\epsilon_{3p3h}} \frac{1}{\epsilon_{3p3h}}
   \langle 0 | \Bigg[ \bigg[ \Big[ \big[ {\bf S}_2, {\mathbf V}_{01} \big],
                                {\mathbf V}_{00}
                          \Big], {\mathbf V}_{00}
                   \bigg], {\mathbf V}_{10} 
            \Bigg] | 2p2h\rangle 
   $
   & (XIV)
   \\ 
   $
   + \frac{1}{2}  \frac{1}{\epsilon_{3p3h}} \frac{1}{\epsilon_{3p3h}}
     \langle 0 | \Bigg[ \bigg[ \Big[ {\bf S}_2, \big[{\bf S}_2, {\mathbf V}_{10} \big]
                            \Big], {\mathbf V}_{00}
                     \bigg], {\mathbf V}_{10}
              \Bigg] | 2p2h\rangle 
   $
   & (XV)
   \\ 
   $
   + \frac{1}{2}  \frac{1}{\epsilon_{4p4h}} \frac{1}{\epsilon_{3p3h}}
     \langle 0 | \Bigg[ \bigg[ \Big[{\bf S}_2, \big[ {\bf S}_2, {\mathbf V}_{00} \big]
                            \Big], {\mathbf V}_{10}
                     \bigg], {\mathbf V}_{10}
              \Bigg] | 2p2h\rangle 
   $
   & (XVI)
   \\ 
   $
   + \frac{1}{2}  \frac{1}{\epsilon_{4p4h}} \frac{1}{\epsilon_{4p4h}}
   \langle 0 | \Bigg[ \bigg[ \Big[ {\bf S}_2, \big[ {\bf S}_2, {\mathbf V}_{00} \big]
                          \Big], {\mathbf V}_{00}
                   \bigg], {\mathbf V}_{20}
            \Bigg] | 2p2h\rangle 
   $
   & (XVII)
   \\ 
   $
   + \frac{1}{2}  \frac{1}{\epsilon_{3p3h}} \frac{1}{\epsilon_{4p4h}}
     \langle 0 | \Bigg[ \bigg[ \Big[ {\bf S}_2, \big[ {\bf S}_2, {\mathbf V}_{10} \big]
                            \Big], {\mathbf V}_{01}
                     \bigg], {\mathbf V}_{20}
              \Bigg] | 2p2h\rangle 
   $
   & (XVIII)
   \\  
   $
   - \frac{1}{\epsilon_{3p3h}} \frac{1}{\epsilon_{3p3h}} \frac{1}{\epsilon_{4p4h}}
   \langle 0 | \Bigg[ \bigg[ \Big[ \big[ {\bf S}_2, {\mathbf V}_{01} \big], 
                                {\mathbf V}_{00} 
                          \Big], {\mathbf V}_{01}
                   \bigg], {\mathbf V}_{20}
            \Bigg] | 2p2h\rangle 
   $
   & (XIX)
\end{tabular}

While we have attempted to be essentially complete in including the
terms $(I) \rightarrow (IV)$, from the others we have selectively included only 
those terms that we expected to be significant. 
If a term is listed but not yet included in our present treatment, 
it will be specifically stated.

\subsection{Effective $ph$-$hp$ Coupled Matrix Element.}
\label{sec:III_a}

We calculate the \nph{2} amplitude as
\begin{eqnarray}
   Z_{p_1p_2,h_1h_2}
   \ = \ 
   - V^{tot}_{p_1p_2,h_1h_2} \ / \ 
   [ \epsilon_{p_1h_1} + \epsilon_{p_2h_2}
     + \Delta\epsilon_{p_1h_1}(\omega=\epsilon_{p_2h_2})
     + \Delta\epsilon_{p_2h_2}(\omega=\epsilon_{p_1h_1}) ]
\end{eqnarray}
or, after angular momentum coupling, 
\begin{eqnarray}
   Z^{\lambda}_{p_1h_1,p_2h_2} 
   \ = \ 
   - \frac{\langle (p_1\bar h_1)_{\lambda}|V^{(tot),\lambda}|(h_2\bar p_2)_{\lambda}\rangle }
          {\epsilon_{p_1h_1} + \epsilon_{p_2h_2} + 
           \Delta\epsilon_{p_1h_1}(\omega=\epsilon_{p_2h_2}) + 
           \Delta\epsilon_{p_2h_2}(\omega=\epsilon_{p_1h_1})}
\end{eqnarray}
This is the only place where we actually use the $\omega$ dependent
contribution to the $ph$ energies.
The total $ph-hp$ matrix element 
$\langle (p_1\bar h_1)_{\lambda}|V^{(tot)}|(h_2\bar p_2)_{\lambda}\rangle $ 
has eight contributions which we label V$_1$-V$_8$.

\begin{enumerate}
   \item{Contribution V$_1$.} 
        The first contribution is the direct matrix element from term (I) 
        \begin{eqnarray}
           \langle (p_1\bar h_1)_{\lambda}|V^{(1),\lambda}|(h_2\bar p_2)_{\lambda}\rangle 
           \ = \ 
           \langle (p_1\bar h_1)_{\lambda}|V^{\lambda}|(h_2\bar p_2)_{\lambda}\rangle 
        \end{eqnarray}
   \item{Contribution V$_2$.}
        The second term is the G-matrix correction generated by term (II)
        \begin{eqnarray}
           V^{(2)}_{p_1p_2,h_1h_2}
           \ = \ 
           \frac{1}{2} \,
           Z_{p_3p_4,h_1h_2} \, V_{p_3p_4,p_1p_2}
           \ + \ 
           \frac{1}{2} \, 
           Z_{p_1p_2,h_3h_4} \, V_{h_3h_4,h_1h_2}
        \end{eqnarray}
        or, 
        \begin{eqnarray}
           (p_1\bar h_1)_{\lambda}|V^{(2),\lambda}|(h_2\bar p_2)_{\lambda}\rangle 
           \ = \ 
           \sum_{K} \ (-)^{K+1} (2K+1) 
                      \wsj{p_1}{h_1}{\lambda}{h_2}{p_2}{K}
                      \langle (p_1p_2)_K|V^{(2),K}|(h_1h_2)_K\rangle 
        \end{eqnarray}     
        where
        \begin{eqnarray}
           \langle (p_1p_2)_K|V^{(2),K}&|(h_1h_2)_K\rangle  
           \ = \ 
           \frac{1}{2} \, 
           \left (
              \langle (p_1p_2)_K | V^K | (p_3p_4)_K\rangle Z^K_{p_3p_4,h_1h_2}
              \ + \
              \langle (h_1h_2)_K | V^K | (h_3h_4)_K\rangle Z^K_{p_1p_2,h_3h_4}
           \right ) 
        \end{eqnarray}
        and
        \begin{eqnarray}
           Z^K_{p_1p_2,h_1h_2}
           \ = \ 
           (-)^{K+1} \ 
           \sum_{\ell} \ (2\ell +1) \
                \wsj{p_1}{p_2}{K}{h_2}{h_1}{\ell} \
                Z^{\ell}_{p_1h_1,p_2h_2}
        \end{eqnarray}
        Note: 
        It is understood that any contribution given in ($pp$) coupling
        in the same way as this term has to be recoupled before it is added to the rest.
   \item{Contribution V$_3$.}
        The third contribution is the collectivity correction also 
        generated from term (II) 
        \begin{eqnarray}
           V^{(3)}_{p_1p_2,h_1h_2}
           \ = \ 
             Z_{p_3h_3,p_1h_1}V_{p_2h_3h_2p_3} 
           + Z_{p_3h_3,p_2h_2}V_{p_1h_3h_1p_3}
           - Z_{p_2h_1,p_3h_3}V_{p_1h_3h_2p_3}
           - Z_{p_3h_3,p_1h_2}V_{p_2h_3h_1p_3}
        \end{eqnarray}
        After angular momentum coupling this becomes
        \begin{eqnarray}
           \langle (p_1\bar h_1)_{\lambda}|V^{(3),\lambda}|(h_2\bar p_2)_{\lambda}\rangle 
           & = & 
           Z^{\lambda}_{p_1h_1,p_3h_3} 
           \langle (p_3\bar h_3)_{\lambda}|V^{eff,\lambda}_{ph}|(p_2\bar h_2)_{\lambda}\rangle  
           \, + \,
           Z^{\lambda}_{p_2h_2,p_3h_3}
           \langle (p_3\bar h_3)_{\lambda}|V^{eff,\lambda}_{ph}|(p_1\bar h_1)_{\lambda}\rangle 
           \nonumber \\  
           && + \ 
           \sum_{\ell} (-)^{\ell+\lambda} (2\ell+1) 
               \wsj{p_1}{h_1}{\lambda}{h_2}{p_2}{\ell}
           \nonumber \\ &&  
           \left (
              Z^{\ell}_{p_1h_2,p_3h_3} 
              \langle (p_3\bar h_3)_{\ell}|V^{eff,\ell}_{ph}|(p_2\bar h_1)_{\ell}\rangle 
              + 
              Z^{\ell}_{p_2h_1,p_3h_3}
              \langle (p_3\bar h_3)_{\ell}|V^{eff,\ell}_{ph}|(p_1\bar h_2)_{\ell}\rangle 
           \right ) 
        \label{eq:2_5}
        \end{eqnarray}         
        with
        \begin{eqnarray}
           \langle (p_3\bar h_3)_{\lambda}|V^{eff,\lambda}_{ph}|(p_2\bar h_2)_{\lambda}\rangle 
           \ = \
           \langle (p_3\bar h_3)_{\lambda}|V^{\lambda}|(p_2\bar h_2)_{\lambda}\rangle  
           \, + \,
           \frac{1}{2} \, 
           Z^{\lambda}_{p_3h_3,p_4h_4}
           \langle (h_4\bar p_4)_{\lambda}|V^{\lambda}|(p_2\bar h_2)_{\lambda}\rangle 
        \label{eq:2_6}
        \end{eqnarray}
        This last correction to the $ph-ph$ matrix element arises from term (III).
   \item{Contribution V$_4$.}
        Additional contributions from term (III) are
        \begin{eqnarray}
           V^{(4)}_{p_1p_2,h_1h_2}
           \ = \ 
           \frac{1}{2} \, 
           Z_{p_1h_3,p_2h_4} Z_{p_3h_1,p_4h_2}
           \langle p_3 \bar h_3 | V | h_4 \bar p_4 \rangle
        \end{eqnarray}
        In angular momentum coupling this term is included as
        \begin{eqnarray}
           \langle (p_1p_2)_K|V^{(4),K}|(h_1h_2)_K\rangle 
           \ = \
           \frac{1}{2} \, 
           Z^K_{p_1p_2,h_3h_4} Z^K_{p_3p_4,h_1h_2} 
           \langle (p_3p_4)_K | V^K | (h_3h_4)_K\rangle 
        \end{eqnarray}
        \emph{Note.} At this point all terms that do not contain an energy denominator
                     have been listed and all two-body terms are included in our treatment.
   \item{Contribution V$_5$.}
        The first energy dependent contributions are generated by term (IV). 
        It is the first term that accounts for the presence of $3p3h$-correlations.
        These terms are discussed in section (\ref{sec:III_a}) below together with higher order terms 
        that can be included using effective matrix elements. 
        We call the sum of all the contributions from section (\ref{sec:III_e}):
        $V^{(5)}_{p_1p_2,h_1h_2}$.
   \item{Contribution V$_6$.}
        These contributions are generated by term (IX) as
        \begin{eqnarray}
           \langle (p_1\bar h_1)_{\lambda}|V^{(6),\lambda}|(h_2\bar p_2)_{\lambda}\rangle 
           \ = \
           - Z^{\lambda}_{p_1h_1,p_6h_6} 
             d(p_6,p_4,\omega=\epsilon_{p_1h_1}+\epsilon_{p_2h_2})
             B^{\lambda}_{p_4h_6,p_5h_5} 
             Z^{\lambda}_{p_5h_5,p_2h_2}
        \end{eqnarray}
        \begin{eqnarray}
           \langle (p_1\bar h_1)_{\lambda}|V^{(6),\lambda}|(h_2\bar p_2)_{\lambda}\rangle 
           \ = \ 
           - Z^{\lambda}_{p_2h_2,p_6h_6}
             d(p_6,p_4,\omega=\epsilon_{p_1h_1}+\epsilon_{p_2h_2})
             B^{\lambda}_{p_4h_6,p_5h_5}
             Z^{\lambda}_{p_5h_5,p_1h_1}
        \end{eqnarray}
        \begin{eqnarray}
           \langle (p_1\bar h_1)_{\lambda}|V^{(6),\lambda}|(h_2\bar p_2)_{\lambda}\rangle 
           \ = \
           - Z^{\lambda}_{p_1h_1,p_6h_6}
             d(h_6,h_4,\omega=\epsilon_{p_1h_1}+\epsilon_{p_2h_2})
             B^{\lambda}_{p_6h_4,p_5h_5} 
             Z^{\lambda}_{p_5h_5,p_2h_2}
        \end{eqnarray}
        \begin{eqnarray}
           \langle (p_1\bar h_1)_{\lambda}|V^{(6),\lambda}|(h_2\bar p_2)_{\lambda}\rangle 
           \ = \
           - Z^{\lambda}_{p_2h_2,p_6h_6} 
             d(h_6,h_4,\omega=\epsilon_{p_1h_1}+\epsilon_{p_2h_2})
             B^{\lambda}_{p_6h_4,p_5h_5}
             Z^{\lambda}_{p_5h_5,p_1h_1}
        \end{eqnarray}
        \begin{eqnarray}
           \langle (p_1\bar h_1)_{\lambda}|V^{(6),\lambda}|(h_2\bar p_2)_{\lambda}\rangle 
           & = & 
           - \sum_{\ell} \, (-)^{\ell+\lambda} (2\ell+1)
                 \wsj{p_1}{h_1}{\lambda}{h_2}{p_2}{\ell}
           \nonumber \\ && \qquad
             Z^{\ell}_{p_1h_2,p_6h_6}
             d(p_6,p_4,\omega=\epsilon_{p_1h_1}+\epsilon_{p_2h_2})
             B^{\ell}_{p_4h_6,p_5h_5}
             Z^{\ell}_{p_5h_5,p_2h_1}
        \end{eqnarray}
        \begin{eqnarray}
           \langle (p_1\bar h_1)_{\lambda}|V^{(6),\lambda}|(h_2\bar p_2)_{\lambda}\rangle 
           & = &
           - \sum_{\ell} \, (-)^{\ell+\lambda} (2\ell+1)
                 \wsj{p_1}{h_1}{\lambda}{h_2}{p_2}{\ell}
           \nonumber \\ && \qquad
             Z^{\ell}_{p_2h_1,p_6h_6}
             d(p_6,p_4,\omega=\epsilon_{p_1h_1}+\epsilon_{p_2h_2})
             B^{\ell}_{p_4h_6,p_5h_5} 
             Z^{\ell}_{p_5h_5,p_1h_2}
        \end{eqnarray}
        \begin{eqnarray}
           \langle (p_1\bar h_1)_{\lambda}|V^{(6),\lambda}|(h_2\bar p_2)_{\lambda}\rangle 
           & = &
           - \sum_{\ell} \, (-)^{\ell+\lambda} (2\ell+1)
                 \wsj{p_1}{h_1}{\lambda}{h_2}{p_2}{\ell}
           \nonumber \\ && \qquad
             Z^{\ell}_{p_1h_2,p_6h_6}
             d(h_6,h_4,\omega=\epsilon_{p_1h_1}+\epsilon_{p_2h_2})
             B^{\ell}_{p_6h_4,p_5h_5}
             Z^{\ell}_{p_5h_5,p_2h_1}
        \end{eqnarray}
        \begin{eqnarray}
           \langle (p_1\bar h_1)_{\lambda}|V^{(6),\lambda}|(h_2\bar p_2)_{\lambda}\rangle 
           & = &
           - \sum_{\ell} \, (-)^{\ell+\lambda} (2\ell+1)
                 \wsj{p_1}{h_1}{\lambda}{h_2}{p_2}{\ell}
           \nonumber \\ && \qquad
             Z^{\ell}_{p_2h_1,p_6h_6}
             d(h_6,h_4,\omega=\epsilon_{p_1h_1}+\epsilon_{p_2h_2})
             B^{\ell}_{p_6h_4,p_5h_5}
             Z^{\ell}_{p_5h_5,p_1h_2}
        \end{eqnarray}
        \begin{eqnarray}
           \langle (p_1p_2)_K|V^{(6),K}|(h_1h_2)_K\rangle 
           & = &
           - \frac{1}{4} \, 
           Z^K_{p_1p_2,h_3h_4}
           d(h_4,h_5,\omega=\epsilon_{p_1h_1}+\epsilon_{p_2h_2})
           B^K_{p_3p_4,h_3,h_5}
           Z^K_{p_3p_4,h_1h_2}
        \end{eqnarray}
        \begin{eqnarray}
           \langle (p_1p_2)_K|V^{(6),K}|(h_1h_2)_K\rangle 
           & = &
           - \frac{1}{4} \,
           Z^K_{p_1p_2,h_3h_4}
           d(h_3,h_5,\omega=\epsilon_{p_1h_1}+\epsilon_{p_2h_2})
           B^K_{p_3p_4,h_5,h_4}
           Z^K_{p_3p_4,h_1h_2}
        \end{eqnarray}
        \begin{eqnarray}
           \langle (p_1p_2)_K|V^{(6),K}|(h_1h_2)_K\rangle 
           & = &
           - \frac{1}{4} \,
           Z^K_{p_1p_2,h_3h_4}
           d(p_4,p_5,\omega=\epsilon_{p_1h_1}+\epsilon_{p_2h_2})
           B_{p_3p_4,h_3h_4}
           Z^K_{p_3p_5,h_1h_2}
        \end{eqnarray}
        \begin{eqnarray}
           \langle (p_1p_2)_K|V^{(6),K}|(h_1h_2)_K\rangle 
           & = &
           - \frac{1}{4} \,
           Z^K_{p_1p_2,h_3h_4}
           d(p_3,p_5,\omega=\epsilon_{p_1h_1}+\epsilon_{p_2h_2})
           B_{p_3p_4,h_3h_4}
           Z^K_{p_5p_4,h_1h_2}
        \end{eqnarray}
   \item{Contribution V$_7$.}
        The contributions $V^{(7)}_{p_1p_2,h_1h_2}$ are the $V^{(2)}$ quenching
        terms and are generated by term (VI) as
        \begin{eqnarray}
           \langle (p_1p_2)_K|V^{(7),K}|(h_1h_2)_K\rangle 
           \ = \
           - \frac{1}{2} \, 
           \sum_K \, 
               d(p_5,p_3,\omega=\epsilon_{p_1h_1}+\epsilon_{p_2h_2})
               Z^K_{p_5p_4,h_1h_2}
               \langle p_3p_4 | V^{K} | p_1p_2 \rangle 
        \end{eqnarray}
        \begin{eqnarray}
           \langle (p_1p_2)_K|V^{(7),K}|(h_1h_2)_K\rangle 
           \ = \
           - \frac{1}{2} \,
           \sum_K \,
               d(h_5,h_3,\omega=\epsilon_{p_1h_1}+\epsilon_{p_2h_2})
               Z^K_{p_1p_2,h_5h_4}
               \langle h_3h_4 | V^{K} | h_1h_2 \rangle
        \end{eqnarray}
        \begin{eqnarray}
           \langle (p_1\bar h_1)_{\lambda}|V^{(7),\lambda}|(h_2\bar p_2)_{\lambda}\rangle 
           \ = \ 
           - d(h3,h4,\omega=\epsilon_{p_1h_1}+\epsilon_{p_2h_2})
             Z^{\lambda}_{p_4h_3,p_2h_2}
             \langle p_4\bar h_4|V^{\lambda}|p_1\bar h_1\rangle 
        \end{eqnarray}
        \begin{eqnarray}
           \langle (p_1\bar h_1)_{\lambda}|V^{(7),\lambda}|(h_2\bar p_2)_{\lambda}\rangle 
           \ = \
           - d(p3,p4,\omega=\epsilon_{p_1h_1}+\epsilon_{p_2h_2})
           Z^{\lambda}_{p_3h_4,p_2h_2}
           \langle p_4\bar h_4|V^{\lambda}|p_1\bar h_1\rangle 
        \end{eqnarray}
        \begin{eqnarray}
           \langle (p_1\bar h_1)_{\lambda}|V^{(7),\lambda}|(h_2\bar p_2)_{\lambda}\rangle 
           \ = \ 
           - d(h3,h4,\omega=\epsilon_{p_1h_1}+\epsilon_{p_2h_2})
           Z^{\lambda}_{p_4h_3,p_1h_1}
           \langle p_4\bar h_4|V^{\lambda}|p_2\bar h_2\rangle 
        \end{eqnarray}
        \begin{eqnarray}
           \langle (p_1\bar h_1)_{\lambda}|V^{(7),\lambda}|(h_2\bar p_2)_{\lambda}\rangle 
           \ = \ 
           - d(p3,p4,\omega=\epsilon_{p_1h_1}+\epsilon_{p_2h_2})
           Z^{\lambda}_{p_3h_4,p_1h_1}
           \langle p_4\bar h_4|V^{\lambda}|p_2\bar h_2\rangle 
        \end{eqnarray}
        \begin{eqnarray}
           \langle (p_1\bar h_1)_{\lambda}|V^{(7),\lambda}|(h_2\bar p_2)_{\lambda}\rangle 
           & = &
           - \sum_{\ell} (-)^{\ell+\lambda} (2\ell+1)
                 \wsj{p_1}{h_1}{\lambda}{h_2}{p_2}{\ell}
           \nonumber \\ && \qquad
             d(h3,h4,\omega=\epsilon_{p_1h_1}+\epsilon_{p_2h_2})
             Z^{\ell}_{p_4h_3,p_1h_2}
             \langle p_4\bar h_4|V^{\ell}|p_2\bar h_1\rangle 
        \end{eqnarray}
        \begin{eqnarray}
           \langle (p_1\bar h_1)_{\lambda}|V^{(7),\lambda}|(h_2\bar p_2)_{\lambda}\rangle 
           & = &
           - \sum_{\ell} (-)^{\ell+\lambda} (2\ell+1)
                 \wsj{p_1}{h_1}{\lambda}{h_2}{p_2}{\ell}
           \nonumber \\ && \qquad
             d(p3,p4,\omega=\epsilon_{p_1h_1}+\epsilon_{p_2h_2})
             Z^{\ell}_{p_3h_4,p_1h_2}
             \langle p_4\bar h_4|V^{\ell}|p_2\bar h_1\rangle 
        \end{eqnarray}
        \begin{eqnarray}
           \langle (p_1\bar h_1)_{\lambda}|V^{(7),\lambda}|(h_2\bar p_2)_{\lambda}\rangle 
           & = &
           - \sum_{\ell} (-)^{\ell+\lambda} (2\ell+1)
                 \wsj{p_1}{h_1}{\lambda}{h_2}{p_2}{\ell}
           \nonumber \\ && \qquad
             d(h3,h4,\omega=\epsilon_{p_1h_1}+\epsilon_{p_2h_2})
             Z^{\ell}_{p_4h_3,p_2h_1}
             \langle p_4\bar h_4|V^{\ell}|p_1\bar h_2\rangle 
        \end{eqnarray}
        \begin{eqnarray}
           \langle (p_1\bar h_1)_{\lambda}|V^{(7),\lambda}|(h_2\bar p_2)_{\lambda}\rangle 
           & = &
           - \sum_{\ell} (-)^{\ell+\lambda} (2\ell+1)
                 \wsj{p_1}{h_1}{\lambda}{h_2}{p_2}{\ell}
           \nonumber \\ && \qquad
             d(p3,p4,\omega=\epsilon_{p_1h_1}+\epsilon_{p_2h_2})
             Z^{\ell}_{p_3h_4,p_2h_1}
             \langle p_4\bar h_4|V^{\ell}|p_1\bar h_2\rangle 
        \end{eqnarray}
   \item{Contribution V$_8$.}
        The last term arises from the fact that the non-symmetric mean field
        has been symmetrized and that the single particle hamiltonian is
        strictly diagonal only for one value of $\omega$. We write
        \begin{eqnarray}
           \langle (p_1\bar h_1)_{\lambda}|V^{(8),\lambda}|(h_2\bar p_2)_{\lambda}\rangle 
           & = &
           \sum_{p_3\neq p_1}
               e(p3,p1,\omega=\epsilon_{p_2h_2})
               Z^{\lambda}_{p_3h_1,p_2h_2}
               \, + \, 
               e(p3,p2,\omega=\epsilon_{p_1h_1})
               Z^{\lambda}_{p_1h_1,p_3h_2}
           \nonumber \\ && 
           \ + \ 
           \sum_{h_1\ne h_3}
               e(h3,h1,\omega=\epsilon_{p_2h_2})
               Z^{\lambda}_{p_1h_3,p_2h_2}
               \, + \, 
               e(h3,h2,\omega=\epsilon_{p_1h_1})
               Z^{\lambda}_{p_1h_1,p_2h_3}
        \end{eqnarray}
        with
        \begin{eqnarray}
           e(p3,p1,\omega)
           & = &
           - \frac{1}{4} \, 
           \frac{2\ell +1}
                {2j_{p_1}+1} \ 
           \sum \ 
                \left (
                   Z^{\ell}_{p_1p_2,h_4h_2}
                   \langle p_3\bar h_4|V^{\ell}|h_2\bar p_2\rangle 
                   \, - \, 
                   Z^{\ell}_{p_3p_2,h_4h_2}
                   \langle p_1\bar h_4|V^{\ell}|h_2\bar p_2\rangle 
                \right )
           \nonumber \\ && 
           - \frac{1}{2} \ 
           \sum_{\ell} \
               \frac{(2\ell+1)}
                    {(2j_{p_1}+1)} \ 
               \frac{\langle p_3\bar p_5 | V^{\ell}| p_6\bar h_6\rangle
                     \langle p_1\bar p_5 | V^{\ell}| p_6\bar h_6\rangle}
                    {\epsilon_{p_6h_6}+\epsilon_{p_5h_1}+\omega}
        \end{eqnarray}
\end{enumerate}

\subsection{Mean Field $pp$ and $hh$ Matrix Elements.}
\label{sec:III_b}

The Schr\"odinger equation for the mean field wave functions are
written in matrix form as
\begin{eqnarray}
   \langle k_1|{\bf H}|k_3\rangle =\langle k_1|{\bf T}|k_3\rangle +\langle k_1|{\bf U}|k_3\rangle 
   \ = \ 
   \delta_{k_1,k_3} \ \epsilon_{k_1}
\end{eqnarray}
where $\epsilon_k$ is the single particle energy of the orbit $k$.
We list here the matrix elements of the effective one-body potential,
and it is understood that the diagonal elements represent the
single particle energies.

The matrix of the single particle potential ${\bf U}$ is computed in
a symmetrized form and in $m$-representation as:
\begin{eqnarray}
   U_{p_1,p_3}
   & = &
   \sum_{h_s} V_{p_1h_s,p_3h_s}
   - \frac{1}{4} 
   \sum_{p_2,h_4,h_2} ( Z_{p_1p_2,h_4h_2}V_{p_3p_2,h_4h_2} +
                        Z_{p_3p_2,h_4h_2}V_{p_1p_2,h_4h_2} )
   \\
   U_{h_1,h_3}
   & = &
   \sum_{h_s}V_{h_sh_1,h_sh_3}
   + \frac{1}{4}
   \sum_{p_4,p_2,h_2} ( Z_{p_4p_2,h_1h_2}V_{p_4p_2,h_3h_2} +
                        Z_{p_4p_2,h_3h_2}V_{p_4p_2,h_1h_2} )
\end{eqnarray}
The second term in these expressions is written in AMC as
\begin{eqnarray}
   - \frac{1}{4} \ \frac{2\ell +1}{2j_{p_1}+1} \ 
   \sum ( Z^{\ell}_{p_1p_2,h_4h_2} \langle p_3\bar h_4|V^{\ell}|h_2\bar p_2\rangle  +
          Z^{\ell}_{p_3p_2,h_4h_2} \langle p_1\bar h_4|V^{\ell}|h_2\bar p_2\rangle  )
\end{eqnarray}
and
\begin{eqnarray}
   + \frac{1}{4} \ \frac{2\ell +1}{2j_{h_1}+1} \ 
   \sum ( Z^{\ell}_{p_4p_2,h_1h_2} \langle p_4\bar h_3|V^{\ell}|h_2\bar p_2\rangle  +
          Z^{\ell}_{p_4p_2,h_3h_2} \langle p_4\bar h_1|V^{\ell}|h_2\bar p_2\rangle  )
\end{eqnarray}
In addition to the contributions in this section we add the
$\omega$ dependent contributions due to corrections for $Z_{3p3h}$
and $Z_{4p4h}$.
Those will be given as contributions to the $ph$-energies.
The $ph$-energies are given as
\begin{eqnarray}
   \epsilon_{p_1,h_1} \ = \
   \epsilon_{p_1}-\epsilon_{h_1}
\end{eqnarray}

\begin{enumerate}
   \item{Term VI.} The following contributions to the single particle potential:
   \begin{enumerate}
      \item
      \begin{eqnarray}
         \Delta U_{p_1h_1}(p_3,p_1,\omega) 
         \ = \
         (-)^{j_{p_5}-j_{p_3}} 
         \sqrt{ \frac{(2j_{p_5}+1)}
                     {(2j_{p_3}+1)} }
         \langle p_1\bar p_3| V^{\ell=0} | p_5\bar p_7 \rangle
         d(p_5,p_7,\omega)
      \end{eqnarray}
      \item
      \begin{eqnarray}
         \Delta U_{p_1h_1}(p_3,p_1,\omega)
         \ = \
         (-)^{j_{p_5}-j_{p_3}}
         \sqrt{ \frac{(2j_{p_5}+1)}
                     {(2j_{p_3}+1)} }
         \langle p_1\bar p_3| V^{\ell=0} | h_5\bar h_7 \rangle
         d(h_5,h_7,\omega)
      \end{eqnarray}
      \item
      \begin{eqnarray}
         \Delta U_{p_1h_1}(h_3,h_1,\omega)
         \ = \
         (-)^{j_{p_5}-j_{h_3}}
         \sqrt{ \frac{(2j_{p_5}+1)}
                     {(2j_{h_3}+1)} }
         \langle h_1\bar h_3| V^{\ell=0} | p_5\bar p_7 \rangle
         d(p_5,p_7,\omega)
      \end{eqnarray}
      \item
      \begin{eqnarray}
         \Delta U_{p_1h_1}(h_3,h_1,\omega)
         \ = \
         (-)^{j_{h_5}-j_{h_3}} 
         \sqrt{ \frac{(2j_{h_5}+1)}
                     {(2j_{h_3}+1)} }
         \langle h_1\bar h_3| V^{\ell=0} | h_5\bar h_7 \rangle
         d(h_5,h_7,\omega)
      \end{eqnarray}
      \item
      \begin{eqnarray}
         \Delta U_{p_1h_1}(p_3,p_1,\omega)
         \ = \ 
         \frac{1}{4}
         \sum_{K} \ 
             \frac{2K+1}
                  {2j_{p_1}+1} \ 
             Z^{K}_{p_1p_5,h_5h_6}
             \langle p_3p_5| V^K| p_7p_8\rangle
             \frac{\langle p_7p_8| V^K| h_5h_6\rangle}
                  {\epsilon_{p_7h_5}+\epsilon_{p_8h_6}+\omega}
      \end{eqnarray}
      By dropping the $\omega$ dependence this is approximated as
      \begin{eqnarray}
         \Delta U_{p_1h_1}(p_3,p_1)
         \ = \
         - \frac{1}{4}
         \sum_{K}
             \frac{2K+1}
                  {2j_{p_1}+1}
             Z^{K}_{p_1p_5,h_5h_6}
             \langle p_3p_5| V^K| p_7p_8\rangle
             Z^K_{p_7p_8,h_5h_6}
      \end{eqnarray}
      \item Similarly we also have
      \begin{eqnarray}
         \Delta U_{p_1h_1}(h_3,h_1)
         \ = \
         - \frac{1}{4}
         \sum_{K}
             \frac{2K+1}
                  {2j_{h_1}+1}
             Z^{K}_{p_5p_6,h_1h_6}
             \langle h_3h_6| V^K| h_7h_8\rangle
             Z^K_{p_5p_6,h_7h_8}
      \end{eqnarray}
      \item
      \begin{eqnarray}
         \Delta U_{p_1h_1}(p_3,p_1,\omega)
         \ = \
         \sum_{\ell} 
             \frac{2\ell+1}{2j_{p_1}+1}
             Z^{\ell}_{p_1h_2,p_5h_5}
             \frac{\langle p_5\bar h_5|V^{\ell}|h_4\bar p_4\rangle }
                  {\epsilon_{p_1h_1}+\epsilon_{p_5h_5}+\epsilon_{p_4h_4}+\omega}
            \langle p_4\bar h_4|V^{\ell}|p_3\bar h_2\rangle 
      \label{eq:3_13}
      \end{eqnarray}
      \item
      \begin{eqnarray}
         \Delta U_{p_1h_1}(h_3,h_1,\omega)
         \ = \
         \sum_{\ell} 
             \frac{2\ell+1}{2j_{p_1}+1}
             Z^{\ell}_{p_3h_1,p_2h_2}
             \frac{\langle p_2\bar h_2|V^{\ell}|h_4\bar p_4\rangle }
                  {\epsilon_{p_1h_1}+\epsilon_{p_2h_2}+\epsilon_{p_4h_4}+\omega}
             \langle p_2\bar h_2|V^{\ell}|p_3\bar h_3\rangle 
      \label{eq:3_14}
      \end{eqnarray}
   \end{enumerate}

   \item{Term IX.} We introduce the notation
        \begin{eqnarray}
           B^{\ell}_{p_1h_1,p_2h_2}
           \ = \
           \langle p_1\bar h_1 | V^{\ell} | h_2\bar p_2\rangle 
        \end{eqnarray}
        The following contributions to the single particle potential are included
   \begin{enumerate}
      \item
      \begin{eqnarray}
         \Delta U_{p_1h_1}(h_a,h_1,\omega)i
         \ = \
         \frac{1}{8} 
         \sum_K
             \frac{2K+1}{2j_{h_1}+1}
             Z^K_{p_3p_4,h_3h_1}
             \frac{B^K_{p_3p_4,h_5h_6}}
                  {\epsilon_{p_3h_5}+\epsilon_{p_4h_6}+\epsilon_{p_1h_1}+\omega}
             Z^K_{p_5p_6,h_5h_6}
             B^K_{p_5p_6,h_3h_a}
      \end{eqnarray}
      \item
      \begin{eqnarray}
         \Delta U_{p_1h_1}(p_a,p_1,\omega)
         \ = \
         \frac{1}{8}
         \sum_K
             \frac{2K+1}{2j_{p_1}+1}
             Z^K_{p_3p_1,h_3h_4}
             \frac{B^K_{p_5p_6,h_3h_4}}
                  {\epsilon_{p_5h_3}+\epsilon_{p_6h_4}+\epsilon_{p_1h_1}+\omega}
             Z^K_{p_5p_6,h_5h_6}
             B^K_{p_3p_a,h_5h_6}      
      \end{eqnarray}
      \item
      \begin{eqnarray}
         \Delta U_{p_1h_1}(p_a,p_1,\omega)
         \ = \
         \frac{1}{2}
         \sum_{\ell}
             \frac{2\ell+1}{2j_{p_1}+1}
             Z^{\ell}_{p_1h_3,p_5h_4}
             B^{\ell}_{p_ah_3,p_4h_4}
             d(p_5,p_4,\omega'=\epsilon_{p_1h_1}+\omega)
      \end{eqnarray}
      \item
      \begin{eqnarray}
         \Delta U_{p_1h_1}(p_a,p_1,\omega)
         \ = \
         \frac{1}{2}
         \sum_{\ell}
             \frac{2\ell+1}{2j_{p_1}+1}
             Z^{\ell}_{p_1h_3,p_4h_5}
             B^{\ell}_{p_ah_3,p_4h_4}
             d(h_5,h_4,\omega'=\epsilon_{p_1h_1}+\omega)
      \end{eqnarray}
      \item
      \begin{eqnarray}
         \Delta U_{p_1h_1}(h_a,h_1,\omega)
         \ = \
         \frac{1}{2}
         \sum_{\ell}
             \frac{2\ell+1}{2j_{h_1}+1}
             Z^{\ell}_{p_3h_1,p_5h_4}
             B^{\ell}_{p_3h_a,p_4h_4}
             d(p_5,p_4,\omega'=\epsilon_{p_1h_1}+\omega)
      \end{eqnarray}
      \item
      \begin{eqnarray}
         \Delta U_{p_1h_1}(h_a,h_1,\omega)
         \ = \
         \frac{1}{2}
         \sum_{\ell}
             \frac{2\ell+1}{2j_{h_1}+1}
             Z^{\ell}_{p_3h_a,p_4h_5}
             B^{\ell}_{p_3h_a,p_4h_4}
             d(h_5,h_4,\omega'=\epsilon_{p_1h_1}+\omega)
      \end{eqnarray}
      \item
      \begin{eqnarray}
         \Delta U_{p_1h_1}(p_a,p_1,\omega)
         \ = \
         - \frac{1}{2}
         Z_{p_3h_3,p_1h_4}
         \frac{B_{p_3h_3,p_5h_5}}
              {\epsilon_{p_3h_3}+\epsilon_{p_5h_5}+\epsilon_{p_1h_1}+\omega}
         Z_{p_5h_5,p_6h_6}
         B_{p_ah_4,p_6h_6}
      \end{eqnarray}
      \item
      \begin{eqnarray}
         \Delta U_{p_1h_1}(h_a,h_1,\omega)
         \ = \
         - \frac{1}{2}
         Z_{p_3h_3,p_4h_1}
         \frac{B_{p_3h_3,p_5h_5}}
              {\epsilon_{p_3h_3}+\epsilon_{p_5h_5}+\epsilon_{p_1h_1}+\omega}
         Z_{p_5h_5,p_6h_6}
         B_{p_4h_a,p_6h_6}
      \end{eqnarray}
      These last two terms are included by using in Eqs.~(\ref{eq:3_13}) 
      and~(\ref{eq:3_14})the effective $ph$-$ph$ matrix element 
      implied by Eq.~(\ref{eq:2_6}).
   \end{enumerate}
\end{enumerate}

\subsection{$\omega$-Dependent Contributions to Single Particle Energies
            Arising from \nph{3}.}
\label{sec:III_c}

These terms are given here as effective $ph$-energies which become $\omega$ dependent. 
They are given as energy matrix $U(h_3,h_1)$ or $U(p_3,p_1)$
as required for the single particle hamiltonian. 
The energies are the diagonal terms, i.e. $h_1=h_3=h$.

\begin{enumerate}
   \item{Terms IV,1 and IV,17.}
   \begin{eqnarray}
      \Delta U_{p_1h_1}(h_3,h_1,\omega)
      \ = \ 
      - \frac{1}{2} 
      \sum_{\ell} 
          \frac{2\ell+1}{2j_{h_1}+1} 
          \frac{\langle h_3\bar h_5 | V^{\ell} | h_6\bar p_6\rangle
                \langle h_1\bar h_5 | V^{\ell} | h_6\bar p_6\rangle}
               {\epsilon_{p_6h_6}+\epsilon_{p_1h_5}+\omega}
   \end{eqnarray}
   \item{Terms IV,13a and IV,13b.}
   \begin{eqnarray}
      \Delta U_{p_1h_1}(p_3,p_1,\omega)
      \ = \ 
      - \frac{1}{2}
      \sum_{\ell} 
          \frac{2\ell+1}{2j_{p_1}+1}
          \frac{\langle p_3\bar p_5 | V^{\ell}| p_6\bar h_6\rangle
                \langle p_1\bar p_5 | V^{\ell}| p_6\bar h_6\rangle}
               {\epsilon_{p_6h_6}+\epsilon_{p_5h_1}+\omega}
   \end{eqnarray}
   Note. The previous two terms are both used with $\omega=\epsilon_{p_2h_2}$.
   \item{Term V,1}
   \begin{eqnarray}
      \Delta U_{p_1h_1}(h_3,h_1,\omega) 
      \ = \ 
      - \frac{1}{2}
      \sum_{\ell} 
          \frac{2\ell+1}{2j_{h_1}+1}
          \frac{\langle h_3\bar h_5 | V^{\ell}| h_4\bar p_4\rangle
                Z^{\ell}_{p_4h_4,p_6h_6}
                \langle h_1\bar h_5 | V^{\ell}| p_6\bar h_6\rangle}
               {\epsilon_{p_6h_6}+\epsilon_{p_1h_5}+\omega}
   \end{eqnarray}
   \begin{eqnarray}
      \Delta U_{p_1h_1}(p_3,p_1,\omega)
      \ = \ 
      - \frac{1}{2}
      \sum_{\ell} 
          \frac{2\ell+1}{2j_{p_1}+1}
          \frac{\langle p_3\bar p_5 | V^{\ell}| p_4\bar h_4\rangle
                Z^{\ell}_{p_4h_4,p_6h_6}
                \langle p_1\bar p_5 | V^{\ell}| h_4\bar p_4\rangle}
               {\epsilon_{p_6h_6}+\epsilon_{p_5h_1}+\omega}
   \end{eqnarray}
   \begin{eqnarray}
      \Delta U_{p_1h_1}(h_3,h_1,\omega)
      \ = \ 
      - \frac{1}{4}
      \sum_K
          \frac{2K+1}{2j_{h_1}+1}
          Z^K_{p_6p_4,h_6h_4}
          \langle p_6p_4| V^K| p_5h_3\rangle
          \frac{\langle h_6h_4| V^K| p_5h_1\rangle}
               {\epsilon_{p_1h_4}+\epsilon_{p_5h_6}+\omega}
   \end{eqnarray}
   \begin{eqnarray}
      \Delta U_{p_1h_1}(p_3,p_1,\omega)
      \ = \ 
      - \frac{1}{4}
      \sum_K
          \frac{2K+1}{2j_{p_1}+1}
          Z^K_{p_6p_4,h_6h_4}
          \langle h_6h_4| V^K| p_3h_5\rangle
          \frac{\langle p_6p_4| V^K| p_1h_5\rangle}
               {\epsilon_{p_4h_1}+\epsilon_{p_6h_5}+\omega}
   \end{eqnarray}
   \item{Term VIII,1}
   \begin{eqnarray}
      \Delta U_{p_1h_1}(p_3,p_1,\omega)
      \ = \
      \frac{1}{2}
      \sum_{\ell} 
      \frac{2\ell+1}{2j_{p_1}+1} 
      \frac{\langle p_4\bar h_4| V^{\ell}| p_6\bar p_3\rangle
            \langle p_4\bar h_4| V^{\ell}| h_5\bar p_5\rangle
            \langle p_5\bar h_5| V^{\ell}| p_6\bar p_1\rangle}
           {(\epsilon_{p_1h_1}+ \epsilon_{p_5h_5}+\epsilon_{p_4h_4}+\omega)
            (\epsilon_{p_6h_1}+ \epsilon_{p_4h_4}+\omega)}
   \end{eqnarray}
   \begin{eqnarray}
      \Delta U_{p_1h_1}(h_3,h_1,\omega)
      \ = \ 
      \frac{1}{2}
      \sum_{\ell} 
      \frac{2\ell+1}{2j_{h_1}+1}
      \frac{\langle p_4\bar h_4| V^{\ell}| h_5\bar h_3\rangle
            \langle p_4\bar h_4| V^{\ell}| h_5\bar p_5\rangle
            \langle p_5\bar h_5| V^{\ell}| h_1\bar h_6\rangle}
           {(\epsilon_{p_1h_1}+ \epsilon_{p_4h_4}+\epsilon_{p_5h_5}+\omega)
            (\epsilon_{p_1h_6}+ \epsilon_{p_4h_4}+\omega)}
   \end{eqnarray}
   \begin{eqnarray}
      \Delta U_{p_1h_1}(h_3,h_1,\omega)
      \ = \
      \frac{1}{4}
      \sum_K 
      \frac{2K+1}{2j_{h_1}+1}
      \frac{\langle h_5h_4| V^K| h_3p_4\rangle
            \langle h_5h_4| V^K| p_5p_6\rangle
            \langle p_5p_6| V^K| h_1p_4\rangle}
           {(\epsilon_{p_1h_1}+ \epsilon_{p_5h_5}+\epsilon_{p_6h_4}+\omega)
            (\epsilon_{p_1h_5}+ \epsilon_{p_4h_4}+\omega)}
   \end{eqnarray}
   \begin{eqnarray}
      \Delta U_{p_1h_1}(p_3,p_1,\omega)
      \ = \
      \frac{1}{4}
      \sum_K 
      \frac{2K+1}{2j_{p_1}+1}
      \frac{\langle p_5p_4| V^K| h_4p_3\rangle
            \langle p_5p_4| V^K| h_5h_6\rangle
            \langle h_5h_6| V^K| h_4p_1\rangle}
           {(\epsilon_{p_1h_1}+\epsilon_{p_5h_5}+\epsilon_{p_4h_6}+\omega)
            (\epsilon_{p_5h_1}+ \epsilon_{p_4h_4}+\omega)}
   \end{eqnarray}
   \item{Term VII,1}
   \begin{eqnarray}
      \Delta U_{p_1h_1}(h_3,h_1,\omega)
      \ = \ 
      \frac{1}{2}
      \sum_{\ell} 
          \frac{2\ell+1}{2j_{h_1}+1} 
          \frac{\langle h_3\bar h_5 | V^{\ell}| h_4\bar p_4\rangle
                \langle p_4\bar h_4|V^{\ell}|p_6\bar h_6\rangle
                \langle h_1\bar h_5 | V^{\ell}| h_6\bar p_6\rangle}
               {(\epsilon_{p_6h_6}+\epsilon_{p_1h_5}+\omega)
                (\epsilon_{p_4h_4}+\epsilon_{p_1h_5}+\omega)}
   \end{eqnarray}
   \begin{eqnarray}
      \Delta U_{p_1h_1}(p_3,p_1,\omega)
      \ = \
      \frac{1}{2}
      \sum_{\ell} 
          \frac{2\ell+1}{2j_{p_1}+1} 
          \frac{\langle p_3\bar p_5 | V^{\ell}| p_4\bar h_4\rangle
                \langle p_4\bar h_4|V^{\ell}|p_6\bar h_6\rangle
                \langle p_1\bar p_5 | V^{\ell}| p_6\bar h_6\rangle}
               {(\epsilon_{p_6h_6}+\epsilon_{p_5h_1}+\omega)
                (\epsilon_{p_4h_4}+\epsilon_{p_5h_1}+\omega)}
   \end{eqnarray}
   \begin{eqnarray}
      \Delta U_{p_1h_1}(h_3,h_1,\omega)
      \ = \
      \frac{1}{4}
      \sum_K 
          \frac{2K+1}{2j_{h_1}+1}
          \frac{\langle h_7h_8| V^K| h_3p_4\rangle
                \langle h_7h_8| V^K| h_5h_6\rangle
                \langle h_5h_6| V^K| h p_4 \rangle}
               {(\epsilon_{p_1h_5}+ \epsilon_{p_4h_6}+\omega)
                (\epsilon_{p_1h_7}+ \epsilon_{p_4h_8}+\omega)}
   \end{eqnarray}
   \begin{eqnarray}
      \Delta U_{p_1h_1}(p_3,p_1,\omega)
      \ = \
      \frac{1}{4}
      \sum_K 
          \frac{2K+1}{2j_{p_1}+1}
          \frac{\langle p_7p_8| V^K| h_4p_3\rangle
                \langle p_7p_8| V^K| p_5p_6\rangle
                \langle p_5p_6| V^K| h_4p_1\rangle}
               {(\epsilon_{p_5h_1}+\epsilon_{p_6h_4}+\omega)
                (\epsilon_{p_7h_1}+\epsilon_{p_8h_4}+\omega)}
   \end{eqnarray}
   \item{Term X,1.}
   \begin{eqnarray}
      \Delta U_{p_1h_1}(h_3,h_1,\omega)
      \ = \
      \frac{1}{2}
      \sum_{\ell} 
          \frac{2\ell+1}{2j_{h_1}+1}
          d(h_8,h_3,\omega) 
          \frac{\langle h_8\bar h_5 | V^{\ell}| h_6\bar p_6\rangle
                \langle h_1\bar h_5 | V^{\ell}| h_6\bar p_6\rangle}
               {\epsilon_{p_6h_6}+\epsilon_{p_1h_5}+\omega}
   \end{eqnarray}
   \begin{eqnarray}
      \Delta U_{p_1h_1}(p_3,p_1,\omega)
      \ = \
      \frac{1}{2}
      \sum_{\ell} 
          \frac{2\ell+1}{2j_{p_1}+1}
          d(p_8,p_3,\omega) 
          \frac{\langle p_8\bar p_5 | V^{\ell}| p_6\bar h_6\rangle
                \langle p_1\bar p_5 | V^{\ell}| p_6\bar h_6\rangle}
               {\epsilon_{p_6h_6}+\epsilon_{p_5h_1}+\omega}
   \end{eqnarray}
   \item{Term XI,1.}
   \begin{eqnarray}
      \Delta U_{p_1h_1}(h_3,h_1,\omega)
      \ = \ 
      \frac{1}{2}
      \sum_{\ell} 
          \frac{2\ell+1}{2j_{h_1}+1} 
          d(h_8,h_6,\omega)
          \frac{\langle h_3\bar h_5 | V^{\ell}| h_6\bar p_6\rangle
                \langle h_1\bar h_5 | V^{\ell}| h_8\bar p_6\rangle}
               {\epsilon_{p_6h_6}+\epsilon_{p_1h_5}+\omega}
   \end{eqnarray}
   \begin{eqnarray}
      \Delta U_{p_1h_1}(h_3,h_1,\omega)
      \ = \
      \frac{1}{2}
      \sum_{\ell} 
          \frac{2\ell+1}{2j_{h_1}+1} 
          d(p_8,p_6,\omega)
          \frac{\langle h_3\bar h_5 | V^{\ell}| h_6\bar p_6\rangle
                \langle h_1\bar h_5 | V^{\ell}| h_6\bar p_8\rangle}
               {\epsilon_{p_6h_6}+\epsilon_{p_1h_5}+\omega}
   \end{eqnarray}
   \begin{eqnarray}
      \Delta U_{p_1h_1}(p_3,p_1,\omega)
      \ = \
      \frac{1}{2}
      \sum_{\ell} 
      \frac{2\ell+1}{2j_{p_1}+1} 
      d(h_8,h_6,\omega)
      \frac{\langle p_3\bar p_5 | V^{\ell}| p_6\bar h_6\rangle
            \langle p_1\bar p_5 | V^{\ell}| p_6\bar h_8\rangle}
           {\epsilon_{p_6h_6}+\epsilon_{p_5h_1}+\omega}
   \end{eqnarray}
   \begin{eqnarray}
      \Delta U_{p_1h_1}(p_3,p_1,\omega)
      \ = \ 
      \frac{1}{2}
      \sum_{\ell} 
      \frac{2\ell+1}{2j_{p_1}+1} 
      d(p_8,p_6,\omega)
      \frac{\langle p_3\bar p_5 | V^{\ell}| p_6\bar h_6\rangle
            \langle p_1\bar p_5 | V^{\ell}| p_8\bar h_6\rangle}
           {\epsilon_{p_6h_6}+\epsilon_{p_5h_1}+\omega}
   \end{eqnarray}
\end{enumerate}

\subsection{Effective $ph$-$ph$ Matrix Element Arising from \nph{3}.}
\label{sec:III_d}

Next we include those terms that can be written similarly to Eq.~(\ref{eq:2_5})
using an effective $ph$-$ph$ matrix element. 
These are generated by Term IV. Similar to Eq.~(\ref{eq:2_5}), 
each term gives rise to four contributions of which we list only one.
The others can be obtained by making the exchange 
($p_1 h_1 \leftrightarrow p_2 h_2$), with either the recoupling 
$h_1 \leftrightarrow h_2$ or $p_1 \leftrightarrow p_2$.
\begin{enumerate}
   \item{Terms IV,2a.}
   \begin{eqnarray}
      - \frac{1}{2} Z_{p_3h_3,p_2h_2}V_{h_5h_6p_1h_3}V_{h_5h_6h_1p_3} / \epsilon
   \end{eqnarray}
   This term can be written as
   \begin{eqnarray}
      \langle (p_1\bar h_1)_{\lambda}|V^{(5),\lambda}|(h_2\bar p_2)_{\lambda}\rangle 
      \ = \
      Z^{\lambda}_{p_3h_3,p_2h_2} A^{eff,\lambda}_{p_1h_1,p_3h_3} / 
                                  (1+\epsilon_{p_2h_2} dA^{eff,\lambda}_{p_1h_1,p_3h_3}/
                                                       A^{eff,\lambda}_{p_1h_1,p_3h_3})
   \label{eq:5_1}                                                    
   \end{eqnarray}
   where $A^{\lambda}$ ($dA^{\lambda}$) are obtained from recoupling the
   $A^{K}$ ($dA^{K}$) according to
   \begin{eqnarray}
      A^{eff,\lambda}(p_1h_1,p_3h_3)
      \ = \
      \sum_{K} (-)^{K+1} (2K+1)
          \wsj{p_1}{h_1}{\lambda}{p_3}{h_3}{K}
          A^{eff,K}(p_1h_3,h_1p_3)
   \label{eq:5_2}                                                    
   \end{eqnarray}
   with 
   \begin{eqnarray}
      A^{eff,K}(p_1h_3,h_1p_3)
      & = &
      - \frac{1}{2}
      \langle p_1h_3| V^{K}| h_5h_6\rangle
      \langle h_1p_3| V^{K}| h_5h_6\rangle / 
      (\epsilon_{p_1h_5}+\epsilon_{p_3h_6}) 
      \\ 
      dA^{eff,K}(p_1h_3,h_1p_3)
      & = &
      - \frac{1}{2}
      \langle p_1h_3| V^{K}| h_5h_6\rangle
      \langle h_1p_3| V^{K}| h_5h_6\rangle /
      (\epsilon_{p_1h_5}+\epsilon_{p_3h_6})^2 
   \end{eqnarray}

   The following corrections arise from term V
   \begin{eqnarray}
      \Delta A^{eff,K}(p_1h_3,h_1p_3)
      & = &
      - \frac{1}{4}
      \langle p_1h_3| V^{K}| h_5h_6\rangle
      \langle h_1p_3| V^{K}| p_7p_8\rangle
      Z^K_{p_7p_8,h_5h_6} /
      (\epsilon_{p_1h_5}+\epsilon_{p_3h_6})
      \\
      \Delta dA^{eff,K}(p_1h_3,h_1p_3)
      & = &
      - \frac{1}{4}
      \langle p_1h_3| V^{K}| h_5h_6\rangle
      \langle h_1p_3| V^{K}| p_7p_8\rangle
      Z^K_{p_7p_8,h_5h_6} /
      (\epsilon_{p_1h_5}+\epsilon_{p_3h_6})^2
   \end{eqnarray}
   and
   \begin{eqnarray}
      \Delta A^{eff,K}(p_1h_3,h_1p_3)
      & = &
      - \frac{1}{2} 
      \sum_\ell \ (-)^{K+1} \ (2 \ell+1) \
          \wsj{p_1}{h_3}{K}{h_6}{h_5}{\ell}
      \nonumber \\ && \hspace{0.5in}
      \langle p_4 \bar h_4| V^{\ell}| h_3 \bar h_6\rangle
      \langle h_1p_3| V^{K}| h_5h_6\rangle
      Z^\ell_{p_4h_4,p_1h_5} /
      (\epsilon_{p_1h_5}+\epsilon_{p_3h_6})
   \label{eq:eq_1}
      \\
      \Delta dA^{eff,K}(p_1h_3,h_1p_3)
      & = &
      - \frac{1}{2} 
      \sum_\ell \ (-)^{K+1} \ (2 \ell+1) \
          \wsj{p_1}{h_3}{K}{h_6}{h_5}{\ell}
      \nonumber \\ && \hspace{0.5in}
      \langle p_4 \bar h_4| V^{\ell}| h_3 \bar h_6\rangle
      \langle h_1p_3| V^{K}| h_5h_6\rangle
      Z^\ell_{p_4h_4,p_1h_5} /
      (\epsilon_{p_1h_5}+\epsilon_{p_3h_6})^2
   \label{eq:eq_2}
   \end{eqnarray}
   and finally,
   \begin{eqnarray}
      \Delta A^{eff,K}(p_1h_3,h_1p_3)
      & = &
      \frac{1}{2}
      \langle p_1h_3| V^{K}| h_5h_6\rangle
      \langle h_1p_3| V^{K}| h_7h_6\rangle d(h_5,h_7) / 
      (\epsilon_{p_1h_5}+\epsilon_{p_3h_6}) 
      \\ 
      \Delta dA^{eff,K}(p_1h_3,h_1p_3)
      & = &
      \frac{1}{2}
      \langle p_1h_3| V^{K}| h_5h_6\rangle
      \langle h_1p_3| V^{K}| h_7h_6\rangle d(h_5,h_7) /
      (\epsilon_{p_1h_5}+\epsilon_{p_3h_6})^2 
   \end{eqnarray}
   Note: Eqs.~(\ref{eq:eq_1}, \ref{eq:eq_2}) are taken into account only 
         when $p_4$ and $h_4$ are both of the same kind (protons or neutrons).

   \item{Terms IV,15a.}
   \begin{eqnarray}
      &&
      - \frac{1}{2} Z_{p_3h_3,p_2h_2}V_{p_3h_1p_5p_6}V_{p_5p_6h_3p_1} / \epsilon
   \end{eqnarray}
   This term can be written as Eq.~(\ref{eq:5_1}) similarly to the previous term, where
   the $A^{\lambda}$ are the recoupled $A^{K}$ according to Eq.~(\ref{eq:5_2}) with    
   \begin{eqnarray}
      A^{eff,K}(p_1h_3,h_1p_3)
      & = & 
      - \frac{1}{2} 
      \langle p_1h_3| V^{K}| p_5p_6\rangle
      \langle h_1p_3| V^{K}| p_5p_6\rangle /
      (\epsilon_{p_5h_1}+\epsilon_{p_6h_3}) 
      \\
      dA^{eff,K}(p_1h_3,h_1p_3)
      & = &
      - \frac{1}{2}
      \langle p_1h_3| V^{K}| p_5p_6\rangle
      \langle h_1p_3| V^{K}| p_5p_6\rangle /
      (\epsilon_{p_5h_1}+\epsilon_{p_6h_3})^2 
   \end{eqnarray}
   with higher order corrections arising from term V, as 
   \begin{eqnarray}
      \Delta A^{eff,K}(p_1h_3,h_1p_3)
      & = &
      - \frac{1}{4}
      \langle p_1h_3| V^{K}| p_5p_6\rangle
      \langle h_1p_3| V^{K}| h_5h_6\rangle
      Z^K_{p_5p_6,h_5h_6} /
      (\epsilon_{p_5h_1}+\epsilon_{p_6h_3})
      \\
      \Delta dA^{eff,K}(p_1h_3,h_1p_3)
      & = &
      - \frac{1}{4}
      \langle p_1h_3| V^{K}| p_5p_6\rangle
      \langle h_1p_3| V^{K}| h_5h_6\rangle
      Z^K_{p_5p_6,h_5h_6} /
      (\epsilon_{p_5h_1}+\epsilon_{p_6h_3})^2
   \end{eqnarray}
   and
   \begin{eqnarray}
      \Delta A^{eff,K}(p_1h_3,h_1p_3)
      & = &
      - \frac{1}{2}  
      \sum_\ell \ (-)^{K+1} \ (2 \ell+1) \
          \wsj{p_1}{h_3}{K}{p_6}{p_5}{\ell}
      \nonumber \\ && \hspace{0.5in}
      \langle p_5 \bar p_1| V^{\ell}| p_7 \bar h_7\rangle
      \langle h_1p_3| V^{K}| p_5p_6\rangle 
      Z^\ell_{p_7h_7,p_6h_3} /
      (\epsilon_{p_5h_1}+\epsilon_{p_6h_3})
   \label{eq:eq_3}
      \\
      \Delta dA^{eff,K}(p_1h_3,h_1p_3)
      & = &
      - \frac{1}{2} 
      \sum_\ell \ (-)^{K+1} \ (2 \ell+1) \
          \wsj{p_1}{h_3}{K}{p_6}{p_5}{\ell}
      \nonumber \\ && \hspace{0.5in}
      \langle p_5 \bar p_1| V^{\ell}| p_7 \bar h_7\rangle
      \langle h_1p_3| V^{K}| p_5p_6\rangle
      Z^\ell_{p_7h_7,p_6h_3} /
      (\epsilon_{p_5h_1}+\epsilon_{p_6h_3})^2
   \label{eq:eq_4}
   \end{eqnarray}
   and
   \begin{eqnarray}
      \Delta A^{eff,K}(p_1h_3,h_1p_3)
      & = & 
      \frac{1}{2} 
      \langle p_1h_3| V^{K}| p_5p_6\rangle
      \langle h_1p_3| V^{K}| p_7p_6\rangle d(p_5,p_7) /
      (\epsilon_{p_5h_1}+\epsilon_{p_6h_3}) 
      \\
      \Delta dA^{eff,K}(p_1h_3,h_1p_3)
      & = &
      \frac{1}{2}
      \langle p_1h_3| V^{K}| p_5p_6\rangle
      \langle h_1p_3| V^{K}| p_7p_6\rangle d(p_5,p_7) /
      (\epsilon_{p_5h_1}+\epsilon_{p_6h_3})^2 
   \end{eqnarray}
   Note: Eqs.~(\ref{eq:eq_3}, \ref{eq:eq_4}) are taken into account only 
         when both $p_7$ and $h_7$ are the same kind of nucleons.

   \item{Terms IV,12a and IV,26b.}
   \begin{eqnarray}
      &&
      Z_{p_3h_3,p_2h_2}V_{p_5p_1h_5p_3}V_{h_3h_5h_1p_5} / \epsilon
   \end{eqnarray}
   This can be written as
   \begin{eqnarray}
      \langle (p_1\bar h_1)_{\lambda}|V^{(5),\lambda}|(h_2\bar p_2)_{\lambda}\rangle 
      \ = \
      Z^{\lambda}_{p_3h_3,p_2h_2} A^{x,\lambda}_{p_1h_1,p_3h_3} /
      (1+\epsilon_{p_2h_2} dA^{x,\lambda}_{p_1h_1,p_3h_3} /
                           A^{x,\lambda}_{p_1h_1,p_3h_3})
   \label{eq:5_9}
   \end{eqnarray}
   where $A^{x,\lambda}$($dA$) are the exchange terms of $W^{\ell}$ ($dW$) 
   according to
   \begin{eqnarray}
      A^{x,\lambda}_{p_1h_1,p_3h_3} 
      \ = \ 
      \sum_{\ell} (-)^{\ell+\lambda} (2\ell+1)
          \wsj{p_1}{h_1}{\lambda}{h_3}{p_3}{\ell}
          W^{\ell}(h_3h_1,p_3p_1)
   \label{eq:5_10}
   \end{eqnarray}
   with
   \begin{eqnarray}
      W^{\ell}(h_3h_1,p_3p_1)
      & = &
      -
      \langle h_3\bar h_1| V^{\ell}| p_5\bar h_5\rangle
      \langle p_5\bar h_5| V^{\ell}| p_3\bar p_1\rangle /
      (\epsilon_{p_5h_5}+\epsilon_{p_1h_3})
      \\
      dW^{\ell}(h_3h_1,p_3p_1)
      & = & 
      -
      \langle h_3\bar h_1| V^{\ell}| p_5\bar h_5\rangle
      \langle p_5\bar h_5| V^{\ell}| p_3\bar p_1\rangle /
      (\epsilon_{p_5h_5}+\epsilon_{p_1h_3})^2
   \end{eqnarray}
   We also have the higher order corrections
   \begin{eqnarray}
      \Delta W^{\ell}(h_3h_1,p_3p_1)
      & = &
      -
      \langle h_1\bar h_3| V^{\ell}| p_6\bar h_6\rangle
      Z_{p_6h_6,p_5h_5}^{\ell}
      \langle p_5\bar h_5| V^{\ell}| p_3\bar p_1\rangle /
      (\epsilon_{p_5h_5}+\epsilon_{p_1h_3})
      \\
      \Delta dW^{\ell}(h_3h_1,p_3p_1)
      & = &
      -
      \langle h_1\bar h_3| V^{\ell}| p_6\bar h_6\rangle
      Z_{p_6h_6,p_5h_5}^{\ell}
      \langle p_5\bar h_5| V^{\ell}| p_3\bar p_1\rangle /
      (\epsilon_{p_5h_5}+\epsilon_{p_1h_3})^2
   \end{eqnarray}
   and
   \begin{eqnarray}
      \Delta W^{\ell}(h_3h_1,p_3p_1)
      & = &
      \langle h_3\bar h_1| V^{\ell}| p_6\bar h_5\rangle
      d(p_5,p_6)
      \langle p_5\bar h_5| V^{\ell}| p_3\bar p_1\rangle /
      (\epsilon_{p_5h_5}+\epsilon_{p_1h_3})
      \\
      \Delta dW^{\ell}(h_3h_1,p_3p_1)
      & = &
      \langle h_3\bar h_1| V^{\ell}| p_6\bar h_5\rangle
      d(p_5,p_6)
      \langle p_5\bar h_5| V^{\ell}| p_3\bar p_1\rangle /
      (\epsilon_{p_5h_5}+\epsilon_{p_1h_3})^2
   \end{eqnarray}
   and
   \begin{eqnarray}
      \Delta W^{\ell}(h_3h_1,p_3p_1)
      & = &
      \langle h_3\bar h_1| V^{\ell}| p_5\bar h_6\rangle
      d(h_5,h_6)
      \langle p_5\bar h_5| V^{\ell}| p_3\bar p_1\rangle /
      (\epsilon_{p_5h_5}+\epsilon_{p_1h_3})
      \\
      \Delta dW^{\ell}(h_3h_1,p_3p_1)
      & = &
      \langle h_3\bar h_1| V^{\ell}| p_5\bar h_6\rangle
      d(h_5,h_6)
      \langle p_5\bar h_5| V^{\ell}| p_3\bar p_1\rangle /
      (\epsilon_{p_5h_5}+\epsilon_{p_1h_3})^2
   \end{eqnarray}

   \item{Terms IV,8b and IV,22a.}
   \begin{eqnarray}
      &&
      Z_{p_3h_3,p_2h_2}V_{p_5h_3h_5h_1}V_{p_1h_5p_3p_5} / \epsilon
   \end{eqnarray}
   This can be written similarly to Eq.~(\ref{eq:5_9}).
   Here the $A^{x,\lambda}$($dA^{x,\lambda}$) are the exchange terms of $W^{\ell}$
   ($dW^{\ell}$) according to Eq.~(\ref{eq:5_10}) where
   \begin{eqnarray}
      W^{\ell}(h_3h_1,p_3p_1)
      & = &
      -
      \langle h_3\bar h_1| V^{\ell}| h_5\bar p_5\rangle
      \langle h_5\bar p_5| V^{\ell}| p_3\bar p_1\rangle /
      (\epsilon_{p_5h_5}+\epsilon_{p_3h_1})
      \\
      dW^{\ell}(h_3h_1,p_3p_1)
      & = & 
      -
      \langle h_3\bar h_1| V^{\ell}| h_5\bar p_5\rangle
      \langle h_5\bar p_5| V^{\ell}| p_3\bar p_1\rangle /
      (\epsilon_{p_5h_5}+\epsilon_{p_3h_1})^2
   \end{eqnarray}
   Here, the higher order corrections are as follows
   \begin{eqnarray}
      W^{\ell}(h_3h_1,p_3p_1)
      & = &
      -
      \langle h_1\bar h_3| V^{\ell}| h_6\bar p_6\rangle
      Z_{p_6h_6,p_5,h_5}^{\ell}
      \langle h_5\bar p_5| V^{\ell}| p_3\bar p_1\rangle /
      (\epsilon_{p_5h_5}+\epsilon_{p_3h_1})
      \\
      dW^{\ell}(h_3h_1,p_3p_1)
      & = &
      -
      \langle h_1\bar h_3| V^{\ell}| h_6\bar p_6\rangle
      Z_{p_6h_6,p_5,h_5}^{\ell}
      \langle h_5\bar p_5| V^{\ell}| p_3\bar p_1\rangle /
      (\epsilon_{p_5h_5}+\epsilon_{p_3h_1})^2
   \end{eqnarray}
   and
   \begin{eqnarray}
      W^{\ell}(h_3h_1,p_3p_1)
      & = &
      \langle h_3\bar h_1| V^{\ell}| h_5\bar p_6\rangle
      d(p_5,p_6)
      \langle h_5\bar p_5| V^{\ell}| p_3\bar p_1\rangle /
      (\epsilon_{p_5h_5}+\epsilon_{p_3h_1})
      \\
      dW^{\ell}(h_3h_1,p_3p_1)
      & = &
      \langle h_3\bar h_1| V^{\ell}| h_5\bar p_6\rangle
      d(p_5,p_6)
      \langle h_5\bar p_5| V^{\ell}| p_3\bar p_1\rangle /
      (\epsilon_{p_5h_5}+\epsilon_{p_3h_1})^2
   \end{eqnarray}
   and
   \begin{eqnarray}
      W^{\ell}(h_3h_1,p_3p_1)
      & = &
      \langle h_3\bar h_1| V^{\ell}| h_6\bar p_5\rangle
      d(h_5,h_6)
      \langle h_5\bar p_5| V^{\ell}| p_3\bar p_1\rangle /
      (\epsilon_{p_5h_5}+\epsilon_{p_3h_1})
      \\
      dW^{\ell}(h_3h_1,p_3p_1)
      & = &
      \langle h_3\bar h_1| V^{\ell}| h_6\bar p_5\rangle
      d(h_5,h_6)
      \langle h_5\bar p_5| V^{\ell}| p_3\bar p_1\rangle /
      (\epsilon_{p_5h_5}+\epsilon_{p_3h_1})^2
   \end{eqnarray}

\end{enumerate}

\subsection{Contributions to V$_5$.}
\label{sec:III_e}

\begin{enumerate}   
   \item{Term IV,20a.}
        \begin{eqnarray}
           &&
           -Z_{p_2h_3,p_4h_4}V_{p_1h_3h_1h_5}V_{p_4h_2h_4h_5}/\epsilon
           \nonumber \\ &&
           \rightarrow \quad
           \langle (p_1\bar h_1)_{\lambda}|V^{(5),\lambda}|(h_2\bar p_2)_{\lambda}\rangle 
           \ = \
           \sum_{\ell} \ (-)^{\ell+\lambda} (2\ell+1)
               \wsj{h_5}{h_3}{\lambda}{p_2}{h_2}{\ell}
           \nonumber \\ && \hspace{2.3in}
               Z^{\ell}_{p_2h_3,p_4h_4}
               \langle p_4\bar h_4 | V^{\ell} | h_5\bar h_2 \rangle
               \frac{\langle p_1\bar h_1| V^{\lambda}| h_5\bar h_3\rangle}
                    {\epsilon_{p_1h_1}+\epsilon_{p_2h_5}+\epsilon_{p_4h_4}}
        \end{eqnarray}
        We can write this as
        \begin{eqnarray}
           \langle (p_1\bar h_1)_{\lambda}|V^{(5),\lambda}|(h_2\bar p_2)_{\lambda}\rangle 
           \ \approx \
           \frac{ZV^{exch,\lambda}(p_2\bar h_2,h_3\bar h_5)
                 \langle p_1\bar h_1| V^{\lambda}| h_5\bar h_3\rangle}
                {(1+\epsilon_{p_1h_1} 
                 dZV^{exch,\lambda}(p_2\bar h_2,h_3\bar h_5) /
                 ZV^{exch,\lambda}(p_2\bar h_2,h_3\bar h_5))}
        \label{eq:6_2}
        \end{eqnarray}
        where the $ZV^{exch,\lambda}$ ($dZV^{exch,\lambda}$) are the
        recoupled $ZV^{\ell}$ ($dZV^{\ell}$) according to
        \begin{eqnarray}
           ZV^{exch,\lambda}(p_2\bar h_2,h_3\bar h_5) 
           \ = \
           \sum_{\ell} (-)^{\ell+\lambda} (2\ell+1)
               \wsj{h_5}{h_3}{\lambda}{p_2}{h_2}{\ell}
               ZV^{\ell}(p_2\bar h_3, h_2\bar h_5)
        \end{eqnarray}
        with
        \begin{eqnarray}
           ZV^{\ell}(p_2\bar h_3, h_2\bar h_5)
           & = &
           Z^{\ell}_{p_2h_3,p_4h_4}
           \langle p_4\bar h_4 | V^{\ell} | h_5\bar h_2\rangle /
           (\epsilon_{p_2h_5}+\epsilon_{p_4h_4})
        \label{eq:eq_141}
           \\
           dZV^{\ell}(p_2\bar h_3, h_2\bar h_5)
           & = &
           Z^{\ell}_{p_2h_3,p_4h_4}
           \langle p_4\bar h_4| V^{\ell} | h_5\bar h_2\rangle /
           (\epsilon_{p_2h_5}+\epsilon_{p_4h_4})^2
        \label{eq:eq_142}
        \end{eqnarray}
        There are additional terms with this symmetry that can be written
        in the same way using effective operators. 
        They are generated by terms V and VIII as well as XI and XIII. 
        Term V corresponds to the following expectation value:
        \begin{eqnarray}
           &&
           -\frac{1}{2} Z_{p_2h_3,p_4h_4} Z_{p_1p_8h_5h_8}
                        V_{p_8h_1h_3h_8} V_{p_4h_2h_4h_5} / \epsilon
        \end{eqnarray}
        This term is incorporated by replacing in (\ref{eq:6_2})
        \begin{eqnarray}
           \langle p_1\bar h_1 | V^{\lambda} | h_5\bar h_3\rangle
           \quad \rightarrow \quad
           \langle p_1\bar h_1 | V^{f_1,\lambda} | h_5\bar h_3\rangle
           & = &
           \langle p_1\bar h_1 | V^{\lambda} | h_5\bar h_3\rangle 
           \, + \,
           \frac{1}{2} Z^{\lambda}_{p_1h_1,p_8h_8}
                       \langle p_8\bar h_8| V^{\lambda}| h_3\bar h_5\rangle
           \nonumber \\ &&
           - \langle p_1\bar h_1 | V^{\lambda} | h_6\bar h_3\rangle d(h_5,h_6)
           - \langle p_1\bar h_1 | V^{\lambda} | h_5\bar h_6\rangle d(h_6,h_3)
        \label{eq:V_f1}
        \end{eqnarray}
        Similarly, we replace in (\ref{eq:eq_141}) and (\ref{eq:eq_142})
        \begin{eqnarray}
           \langle p_4\bar h_4 | V^{\ell} | h_5\bar h_2\rangle
           \quad \rightarrow \quad
           \langle p_4\bar h_4 | V^{f_2,\ell} | h_5\bar h_2\rangle
           & = &
           \langle p_4\bar h_4 | V^{\ell} | h_5\bar h_2\rangle
           \, + \,
           Z^{\ell}_{p_5h_5,p_4h_4}
           \langle p_5\bar h_5| V^{\ell}| h_2\bar h_5\rangle
           \nonumber \\ &&
           - \langle p_6\bar h_4 | V^{\ell} | h_5\bar h_2\rangle d(p_4,p_6)
           - \langle p_4\bar h_6 | V^{\ell} | h_5\bar h_2\rangle d(h_4,h_6)
        \end{eqnarray}

   \item{Term IV,16b.}
        \begin{eqnarray}
           &&
           -Z_{p_3h_3,p_4h_2}V_{p_1p_5h_1p_4}V_{p_3p_5h_3p_2} / \epsilon
        \end{eqnarray}
        gives
        \begin{eqnarray}
           \langle (p_1\bar h_1)_{\lambda}|V^{(5),\lambda}|(h_2\bar p_2)_{\lambda}\rangle 
           \ \approx \
           \frac{ZV^{exch,\lambda}(p_2\bar h_2,p_5\bar p_4)
                 \langle p_1\bar h_1| V^{\lambda}| p_4\bar p_5\rangle }
                {1+\epsilon_{p_1h_1} dZV^{exch,\lambda}(p_2\bar h_2,p_4\bar p_5) /
                                     ZV^{exch,\lambda}(p_2\bar h_2,p_5\bar p_4) }
        \label{eq:6_2like}
        \end{eqnarray}
        where the $ZV^{exch,\lambda}$ ($dZV^{exch,\lambda}$) are the
        recoupled $ZV^{\ell}$ ($dZV^{\ell}$) according to:
        \begin{eqnarray}
           ZV^{exch,\lambda}(p_2\bar h_2,p_5\bar p_4)
           \ = \
           \sum_{\ell} (-)^{\ell+\lambda} (2\ell+1)
           \wsj{p_5}{p_4}{\lambda}{p_2}{h_2}{\ell}
           ZV^{\ell}(p_4\bar h_2,p_2\bar p_5)
        \end{eqnarray}
        with
        \begin{eqnarray}
           ZV^{\ell}(p_4\bar h_2,p_2\bar p_5)
           & = &
           \frac{Z^{\ell}_{p_3h_3,p_4h_2}
                 \langle p_3\bar h_3 | V^{\ell} | p_2\bar p_5\rangle}
                {\epsilon_{p_5h_2}+\epsilon_{p_3h_3}}
        \label{eq:eq_148a}
           \\
           dZV^{\ell}(p_4\bar h_2,p_2\bar p_5)
           & = &
           \frac{Z^{\ell}_{p_3h_3,p_4h_2}
                 \langle p_3\bar h_3 | V^{\ell} | p_2\bar p_5\rangle}
                {(\epsilon_{p_5h_2}+\epsilon_{p_3h_3})^2}
        \label{eq:eq_148b}
        \end{eqnarray}
        As before, additional terms are incorporated using the substitution
        in (\ref{eq:6_2like})
        \begin{eqnarray}
           \langle p_1\bar h_1 | V^{\lambda} | p_4\bar p_5\rangle
           \quad \rightarrow \quad
           \langle p_1\bar h_1 | V^{f_1,\lambda} | p_4\bar p_5\rangle
           & = &
           \langle p_1\bar h_1 | V^{\lambda} | p_4\bar p_5\rangle 
           \, + \,
           \frac{1}{2} Z^{\lambda}_{p_1h_1,p_8h_8}
                       \langle p_8\bar h_8| V^{\lambda}| p_5\bar p_4\rangle
           \nonumber \\ &&
           - \langle p_1\bar h_1 | V^{\lambda} | p_8\bar p_5\rangle d(p_8,p_4)
           - \langle p_1\bar h_1 | V^{\lambda} | p_4\bar p_8\rangle d(p_8,p_5)
        \end{eqnarray}
        and also, we replace in (\ref{eq:eq_148a}) and (\ref{eq:eq_148b})
        \begin{eqnarray}
           \langle p_3\bar h_3 | V^{\ell} | p_2\bar p_5\rangle
           \quad \rightarrow \quad
           \langle p_3\bar h_3 | V^{f_2,\ell} | p_2\bar p_5\rangle
           & = &
           \langle p_3\bar h_3 | V^{\ell} | p_2\bar p_5\rangle
           \, + \,
           Z^{\ell}_{p_6h_6,p_3h_3}
           \langle p_6\bar h_6| V^{\ell}| p_5\bar p_2\rangle
           \nonumber \\ &&
           - \langle p_6\bar h_3 | V^{\ell} | p_2\bar p_5\rangle d(p_6,p_3)
           - \langle p_3\bar h_6 | V^{\ell} | p_2\bar p_5\rangle d(h_3,h_6)
        \end{eqnarray}

   \item{Terms IV, 4b and 28a.}
        \begin{eqnarray}
           &&
           -Z_{p_1h_3,p_4h_4}V_{p_2h_3h_2h_5}V_{p_4h_1h_4h_5} / \epsilon
           \\ &&
           -Z_{p_3h_3,p_4h_1}V_{p_2p_5h_2p_4}V_{p_3p_5h_3p_1} / \epsilon
        \end{eqnarray}
        These are the previous two terms with $p_1h_1$ and $p_2h_2$ exchanged.
        We incorporate them as such with all the corresponding higher order
        corrections included.

   \item{Term IV, 20b, 4a, 16a, and 28b.}
        These terms are the exchange terms to the previous four terms with
        $p_1$ and $p_2$ exchanged. 
        They are computed from the previous expressions with the additional recoupling:
        \begin{eqnarray}
           \langle p_1\bar h_1|V^{(5),\lambda}|h_2\bar p_2\rangle 
           \ = \
           \sum_{\ell} (-)^{\ell+\lambda} (2\ell+1)
               \wsj{p_1}{h_1}{\lambda}{h_2}{p_2}{\ell}
               \langle (p_2\bar h_1)_{\ell}|V^{(5),\ell}|(h_2\bar p_1)_{\ell}\rangle 
        \label{eq:6_20}
        \end{eqnarray}

   \item{Term IV, 6a.}
        \begin{eqnarray}
           &&
           \frac{1}{2}
           Z_{p_3h_3,p_1h_4}V_{p_3p_6h_3p_2}V_{h_2h_1p_6h_4} / \epsilon
        \end{eqnarray}
        gives
        \begin{eqnarray}
           \Delta B^K(p_1p_2,h_1h_2)
           \ \approx \
           \langle h_1h_2|V^K|p_6h_4\rangle
           \frac{ZV^K(p_1p_2,p_6h_4)}
                {1 + (\epsilon_{p_2h_2}+\epsilon_{h_1}-\epsilon_{h_4})
                     dZV^K(p_1p_2,p_6h_4)}
        \label{eq:eq_140}
        \end{eqnarray}
        where the $ZV^K$($dZV^K$) are the recoupled quantities from term (20)
        according to
        \begin{eqnarray}
           ZV^K(p_1p_2,p_6h_4)
           \ = \
           \sum_{\ell} (-)^{K+1} \ (2\ell+1) \
               \wsj{p_1}{p_2}{K}{h_4}{p_6}{\ell} \
               ZV^{\ell}(p_2h_4,p_6p_1)
        \end{eqnarray}
        Higher order corrections are included by replacing in (\ref{eq:eq_140})
        \begin{eqnarray}
           \langle h_1h_2 | V^{K} | p_6h_4 \rangle
           \quad \rightarrow \quad
           \langle h_1h_2 | V^{f_3,K} | p_6h_4 \rangle
           & = &
           \langle h_1h_2 | V^{K} | p_6h_4 \rangle
           \, + \,
           \frac{1}{2}
           Z^{K}_{p_3p_4,h_1h_2}
           \langle p_3p_4| V^{K}| p_6h_4\rangle
        \end{eqnarray}

   \item{Term IV, 6b.}
        \begin{eqnarray}
           &&
           \frac{1}{2}
           Z_{p_3h_3,p_1h_4}V_{p_3p_6h_3p_2}V_{h_2h_1p_6h_4} / \epsilon
        \end{eqnarray}
        gives
        \begin{eqnarray}
           \Delta B^K(p_1p_2,h_1h_2)
           \ \approx \
           \langle h_2h_1|V^K|p_6h_4\rangle
           \frac{ZV^K(p_2p_1,p_6h_4)}
                {1 + (\epsilon_{p_1h_1}+\epsilon_{h_2}-\epsilon_{h_4})
                     dZV^K(p_2p_1,p_6h_4)}
        \end{eqnarray}
        For the $ph$ coupled contribution it can be written as the transpose
        of term (6a).

   \item{Term IV, 11.}
        \begin{eqnarray}
           &&
           \frac{1}{2}
           Z_{p_3h_2,p_4h_4}V_{p_1p_2h_6p_3}V_{h_6h_4h_1p_4} / \epsilon
        \end{eqnarray}
        gives
        \begin{eqnarray}
           \Delta B^K(p_1p_2,h_1h_2)
           \ \approx \
           \frac{\langle p_1p_2|V^K|h_6p_3\rangle
                 ZV^K(h_1h_2,h_6p_3)}
                {1 + (\epsilon_{p_2h_2}+\epsilon_{p_1}-\epsilon_{p_3})
                     dZV^K(h_1h_2,h_6p_3)/ZV^K(h_1h_2,h_6p_3)}
        \label{eq:eq_145}
        \end{eqnarray}
        where the $ZV^K(h_1h_2,h_6p_3)$ are the $pp$-coupled $ZV$ from
        term (16) according to
        \begin{eqnarray}
           ZV^K(h_1h_2,h_6p_3)
           \ = \
           \sum_{\ell} (-)^{K+1} \, (2\ell+1) \
               \wsj{p_3}{h_6}{K}{h_1}{h_2}{\ell} \
               ZV^{\ell}(p_3\bar h_2,h_6\bar h_1)
        \end{eqnarray}
        Higher order corrections are included by replacing in (\ref{eq:eq_145})
        \begin{eqnarray}
           \langle p_1p_2 | V^{K} | h_6p_3 \rangle
           \quad \rightarrow \quad
           \langle p_1p_2 | V^{f_3,K} | h_6p_3 \rangle
           & = &
           \langle p_1p_2 | V^{K} | h_6p_3 \rangle
           \, + \,
           \frac{1}{2}
           Z^{K}_{p_1p_2,h_3h_4}
           \langle h_3h_4| V^{K}| h_6p_3 \rangle
        \end{eqnarray}

   \item{Term IV, 25.}
        This is the same as the previous term with $h_1$ and $h_2$ interchanged.

   \item{Term IV,7a.}
        \begin{eqnarray}
           &&
           -\frac{1}{2} Z_{p_3h_3,p_4h_1}V_{p_2h_3h_2h_6}V_{p_3p_4p_1h_6} / \epsilon
        \end{eqnarray}
        gives
        \begin{eqnarray}
           \langle (p_1\bar h_1)_{\lambda}|V^{(5),\lambda}|(h_2\bar p_2)_{\lambda}\rangle 
           & = & 
           \sum_K \ \frac{1}{2} \, 
                  \frac{Z^K_{p_3p_4,h_1h_3} 
                        \langle p_3p_4 | V^K | p_1h_6\rangle}
                       {\epsilon_{p_4h_1}+\epsilon_{p_2h_2}+\epsilon_{p_3h_6}}
           \nonumber \\ && \qquad
                  (-)^{K+1} (2K+1)
                  \wsj{p_1}{h_1}{\lambda}{h_3}{h_6}{K}
                  \langle p_2\bar h_2 | V^{\lambda} | h_6\bar h_3\rangle
        \label{eq:6_10}
        \end{eqnarray}
        Higher order corrections are included by using in (\ref{eq:6_10})
        the substitution (\ref{eq:V_f1})
        \begin{eqnarray}
           \langle p_2\bar h_2 | V^{\lambda} | h_6\bar h_3 \rangle
           \quad \rightarrow \quad
           \langle p_2\bar h_2 | V^{f_1,\lambda} | h_6\bar h_3 \rangle
        \end{eqnarray}
        We approximate this by
        \begin{eqnarray}
           \langle (p_1\bar h_1)_{\lambda}|V^{(5),\lambda}|(h_2\bar p_2)_{\lambda}\rangle 
           \ \approx \
           \frac{ZV^{r,\lambda}(p_1\bar h_1,h_3\bar h_5)
                 \langle p_2\bar h_2| V^{\lambda}| h_5\bar h_3\rangle}
                {(1+\epsilon_{p_2h_2} dZV^{r,\lambda}(p_1\bar h_1,h_3\bar h_5) /
                                      ZV^{r,\lambda}(p_1\bar h_1,h_3\bar h_5))}
        \end{eqnarray}
        Here the $ZV^{r,\lambda}$ ($dZV^{r,\lambda}$) are recoupled from
        $pp$ coupling via
        \begin{eqnarray}
           ZV^{r,\lambda}(p_2\bar h_2,h_3\bar h_5))
           \ = \
           (-)^{K+1} \, (2K+1) \ 
           \wsj{p_1}{h_1}{\lambda}{h_3}{h_6}{K} \
           ZV^K(p_1h_6,h_1h_3)
        \end{eqnarray}
        with
        \begin{eqnarray}
           ZV^K(p_1h_6,h_1h_3)
           & = & 
           \frac{Z^K_{p_3p_4,h_1h_3}
                 \langle p_3p_4 | V^K | p_1h_6\rangle}
                {\epsilon_{p_4h_1}+\epsilon_{p_3h_6}}
           \\
           dZV^K(p_1h_6,h_1h_3)
           & = &
           \frac{Z^K_{p_3p_4,h_1h_3}
                 \langle p_3p_4 | V^K | p_1h_6\rangle}
                {(\epsilon_{p_4h_1}+\epsilon_{p_3h_6})^2}
        \end{eqnarray}

   \item{Term IV, 10b.}
        \begin{eqnarray}
           &&
           - \frac{1}{2} Z_{p_3h_3,p_1h_4}V_{p_7p_2p_3h_2}V_{h_3h_4h_1p_7} / \epsilon
        \end{eqnarray}
        This term is written similarly to Eq.~(\ref{eq:6_10}) as
        \begin{eqnarray}
           \langle (p_1\bar h_1)_{\lambda}|V^{(5),\lambda}|(h_2\bar p_2)_{\lambda}\rangle 
           \ \approx \
           \frac{ZV^{r,\lambda}(p_1\bar h_1,p_7\bar p_3)
                 \langle p_2\bar h_2| V^{\lambda}| p_3\bar p_7\rangle}
                {1+\epsilon_{p_2h_2} dZV^{r,\lambda}(p_1\bar h_1,p_7\bar p_3)/
                                     ZV^{r,\lambda}(p_1\bar h_1,p_7\bar p_3))}
        \label{eq:3_154}
        \end{eqnarray}
        Here the $ZV^{r,\lambda}$ ($dZV^{r,\lambda}$) are recoupled from
        $pp$ coupling via
        \begin{eqnarray}
           ZV^{r,\lambda}(p_1\bar h_1,p_7\bar p_3)
           \ = \ 
           \sum_K \ \frac{1}{2} 
                  (-)^{K+1} \, (2K+1) \
                  \wsj{p_1}{h_1}{\lambda}{p_7}{p_3}{K} \
                  ZV^K(p_3\bar p_1,h_1\bar p_7)
        \end{eqnarray}
        where we have
        \begin{eqnarray}
           ZV^K(p_3\bar p_1,h_1\bar p_7)
           & = & 
           \frac{1}{2} \,
           \frac{Z^K_{p_3p_1,h_3h_4}
                 \langle h_3h_4 | V^K | h_1p_7 \rangle}
                {\epsilon_{p_1h_3}+\epsilon_{p_7h_4}}
           \\
           dZV^K(p_3\bar p_1,h_1\bar p_7)
           & = & 
           \frac{1}{2} \,
           \frac{Z^K_{p_3p_1,h_3h_4}
                 \langle h_3h_4 | V^K | h_1p_7\rangle}
                {(\epsilon_{p_1h_3}+\epsilon_{p_7h_4})^2}
        \end{eqnarray}
        Higher order corrections are obtained by using in (\ref{eq:3_154})
        the substitution (\ref{eq:V_f1})
        \begin{eqnarray}
           \langle p_2\bar h_2 | V^{\lambda} | p_3\bar p_7 \rangle
           \quad \rightarrow \quad
           \langle p_2\bar h_2 | V^{f_1,\lambda} | p_3\bar p_7 \rangle
        \end{eqnarray}

   \item{Term IV, 21b and 10a.}
        These are obtained from the previous two terms by interchanging
        $p_1h_1$ and $p_2h_2$, and they are computed that way.

   \item{Term IV, 21a, 7b, 24a, and 24b.}
        These terms are the exchange terms to the previous four terms with
        $p_1$ and $p_2$ exchanged. 
        They are computed from the previous
        expressions with the additional recoupling given by Eq.~(\ref{eq:6_20}).

   \item{Term IV, 19.}
        \begin{eqnarray}
           &&
           -Z_{p_1h_3,p_2h_4}V_{p_5h_3h_5h_1}V_{h_4h_5h_2p_5} / \epsilon
        \end{eqnarray}
        gives
        \begin{eqnarray}
           \Delta B^K(p_1p_2,h_1h_2)
           \ = \ 
           \frac{Z^K_{p_1p_2,h_3,h_4}
                 W^K_{h_1h_2,h_3h_4}}
                {1 + (\epsilon_{p_1}+\epsilon_{p_2}) 
                     dW^K_{h_1h_2,h_3h_4}/W^K_{h_1h_2,h_3h_4}}
        \end{eqnarray}
        with
        \begin{eqnarray}
           (d)W^K_{h_1h_2,h_3h_4}
           \ = \ 
           \sum_{\ell} (-)^{K+1} \, (2\ell+1) \
               \wsj{h_1}{h_2}{K}{h_3}{h_4}{\ell} \
               (d)W^{\ell}_{h_1\bar h_3,h_4\bar h_2}
        \end{eqnarray}
        and
        \begin{eqnarray}
           W^{\ell}_{h_1\bar h_3,h_4\bar h_2}
           & = & 
           - \, \langle h_1\bar h_3| V^{\ell}| p_5\bar h_5\rangle
                \langle h_4\bar h_2| V^{\ell}| p_5\bar h_5\rangle / 
                (\epsilon_{h_4}+\epsilon_{h_1}+\epsilon_{p_5h_5})
           \\
           dW^{\ell}_{h_1\bar h_3,h_4\bar h_2}
           & = &
           - \, \langle h_1\bar h_3| V^{\ell}| p_5\bar h_5\rangle
                \langle h_4\bar h_2| V^{\ell}| p_5\bar h_5\rangle /
                (\epsilon_{h_4}+\epsilon_{h_1}+\epsilon_{p_5h_5})^2
        \end{eqnarray}

   \item{Term IV, 3.}
        \begin{eqnarray}
           &&
           -Z_{p_2h_3,p_1h_4}V_{p_5h_3h_5h_2}V_{h_4h_5h_1p_5} / \epsilon
        \end{eqnarray}
        This term arises from the exchange ($p_1h_1$) with ($p_2h_2$).
        It thus results in the transpose of term (19). 
        Combining terms (3) and (19) gives
        \begin{eqnarray}
           &&
           \Delta B^K(p_1p_2,h_1h_2)
           \ = \ 
           Z^K_{p_1p_2,h_3,h_4}
           \\ && \times \ 
           \left [ 
              \frac{W^K_{h_1h_2,h_3h_4}}
                   {1 + (\epsilon_{p_1}+\epsilon_{p_2}) 
                        dW^K_{h_1h_2,h_3h_4}/W^K_{h_1h_2,h_3h_4}}
               \, + \, 
               (-)^K \, 
               \frac{W^K_{h_2h_1,h_3h_4}}
                    {1 + (\epsilon_{p_1}+\epsilon_{p_2})
                         dW^K_{h_2h_1,h_3h_4}/W^K_{h_2h_1,h_3h_4}}
           \right ]
        \end{eqnarray}
        Here the sum is not restricted and goes over $h_3 \rangle  h_4$ as well
        as $h_4 \rangle  h_3$. 
        By using a restricted sum (only $h_3 \rangle  h_4$, we write
        \begin{eqnarray}
           &&
           \Delta B^K(p_1p_2,h_1h_2)
           \ = \
           Z^K_{p_1p_2,h_3,h_4}
           \\ && \times \
           \left [
              \frac{W^K_{h_1h_2,h_3h_4}}
                   {1 + (\epsilon_{p_1}+\epsilon_{p_2})
                        dW^K_{h_1h_2,h_3h_4}/W^K_{h_1h_2,h_3h_4}}
              \, + \,
              (-)^K
              \frac{W^K_{h_2h_1,h_3h_4}}
                   {1 + (\epsilon_{p_1}+\epsilon_{p_2})
                        dW^K_{h_2h_1,h_3h_4}/W^K_{h_2h_1,h_3h_4}}
           \right .
           \\ && \quad
           \left .
              \, + \,
              \frac{W^K_{h_2h_1,h_4h_3}}
                   {1 + (\epsilon_{p_1}+\epsilon_{p_2})
                        dW^K_{h_2h_1,h_4h_3}/W^K_{h_2h_1,h_4h_3}}
              \, + \,
              (-)^K
              \frac{W^K_{h_1h_2,h_4h_3}}
                   {1 + (\epsilon_{p_1}+\epsilon_{p_2})
                        dW^K_{h_1h_2,h_4h_3}/W^K_{h_1h_2,h_4h_3}}
           \right ]
        \end{eqnarray}

   \item{Term IV, 14a.}
        \begin{eqnarray}
           &&
           -Z_{p_3p_4,h_2h_1}V_{p_3h_5p_2p_5}V_{p_4p_5p_1h_5} / \epsilon
        \end{eqnarray}
        gives
        \begin{eqnarray}
           \Delta B^K(p_1p_2,h_1h_2)
           \ = \
           \frac{Z^K_{p_3p_4,h_1,h_2}
                 W^K_{p_1p_2,p_3p_4}}
                {1 + (\epsilon_{h_1}+\epsilon_{h_2}) 
                     dW^K_{p_1p_2,p_3p_4}/W^K_{p_1p_2,p_3p_4}}
        \end{eqnarray}
        This is similar to terms (3) and (19) except that
        \begin{eqnarray}
           W^{\ell}_{p_1\bar p_3,p_4\bar p_2}
           & = &
           - \langle p_3\bar p_1 | V^{\ell} | p_5\bar h_5 \rangle
             \langle p_2\bar p_4 | V^{\ell} | p_5\bar h_5\rangle
             \, / \,
             (\epsilon_{p_4}+\epsilon_{p_1}+\epsilon_{p_5h_5})
           \\
           dW^{\ell}_{p_1\bar p_3,p_4\bar p_2}
           & = &
           - \langle p_3\bar p_1| V^{\ell}| p_5\bar h_5\rangle
             \langle p_2\bar p_4| V^{\ell}| p_5\bar h_5\rangle
             \, / \,
             (\epsilon_{p_4}+\epsilon_{p_1}+\epsilon_{p_5h_5})^2
        \end{eqnarray}

   \item{Term IV, 14b.}
        \begin{eqnarray}
           &&
           -Z_{p_3p_4,h_1h_2}V_{p_3h_5p_1p_5}V_{p_4p_5p_2h_5} / \epsilon
        \end{eqnarray}
        This term again arises from the exchange ($p_1h_1$) with ($p_2h_2$).
        It thus results in the transpose of term (14a).
        We combine the two terms using restricted sums as
        \begin{eqnarray}
           &&
           \Delta B^K(p_1p_2,h_1h_2)
           \ = \ Z^K_{p_3p_4,h_1,h_2}
           \\ && \times \
           \left [ 
              \frac{W^K_{p_1p_2,p_3p_4}}
                   {1 + (\epsilon_{h_1}+\epsilon_{h_2})
                        dW^K_{p_1p_2,p_3p_4} / W^K_{p_1p_2,p_3p_4}}
              \, + \, 
              (-)^K 
              \frac{W^K_{p_2p_1,p_3p_4}}
                   {1 + (\epsilon_{h_1}+\epsilon_{h_2})
                        dW^K_{p_2p_1,p_3p_4} / W^K_{p_2p_1,p_3p_4}}
           \right .
           \nonumber \\ && \quad
           \left .
              \frac{W^K_{p_2p_1,p_4p_3}}
                   {1 + (\epsilon_{h_1}+\epsilon_{h_2})
                   dW^K_{p_2p_1,p_4p_3} / W^K_{p_2p_1,p_4p_3}}
              \, + \,
              (-)^K
              \frac{W^K_{p_1p_2,p_4p_3}}
                   {1 + (\epsilon_{h_1}+\epsilon_{h_2})
                        dW^K_{p_1p_2,p_4p_3} / W^K_{p_1p_2,p_4p_3}}
           \right ]
        \end{eqnarray}
\end{enumerate}   

\hspace{0.2in}

\emph{Note.} Contributions from \nph{3} have been included in full except for 
     the last two terms (14a and 14b), which are listed for completeness.

\end{document}